\documentclass[acmlarge,authoryear]{acmart}
\usepackage{etex} 
\usepackage[utf8x]{inputenc}
\usepackage{xspace}
\usepackage{proof}
\usepackage{amsfonts}
\usepackage{amsthm}
\usepackage{amssymb}
\usepackage{amsmath}
\usepackage{stmaryrd}
\usepackage{fancyvrb}
\usepackage{paralist}
\usepackage{graphicx}
\usepackage{fancyhdr}
\usepackage{placeins} 
\usepackage{soul}
\usepackage{fancyvrb}
\usepackage{enumerate}
\usepackage[all]{xy}
\usepackage{url}
\usepackage{mathtools}
\usepackage{hyperref}
\usepackage{bussproofs}
\usepackage{mathpartir}
\usepackage{bm}
\usepackage{float}
\usepackage{xcolor}
\usepackage[cal=boondoxo]{mathalfa}
\usepackage{adjustbox}
\usepackage{pifont} 
\usepackage{arydshln}
\usepackage[xcolor]{changebar}
\cbcolor{red}
\DeclareFontFamily{U}{matha}{\hyphenchar\font45}
\DeclareFontShape{U}{matha}{m}{n}{
      <5> <6> <7> <8> <9> <10> gen * matha
      <10.95> matha10 <12> <14.4> <17.28> <20.74> <24.88> matha12
      }{}
\DeclareSymbolFont{matha}{U}{matha}{m}{n}

\DeclareMathSymbol{\twoPrime}{3}{matha}{"32}
\DeclareMathSymbol{\threePrime}{3}{matha}{"33}
\DeclareMathSymbol{\fourPrime}{3}{matha}{"34}

\hypersetup{
    colorlinks=true,
    citecolor=blue,
    linkcolor=blue,
    urlcolor=blue
}

\allowdisplaybreaks

\makeatletter
\newcommand*{\etal}{%
    \@ifnextchar{.}%
        {et al}%
        {et al.\@\xspace}%
}
\makeatother
\newcommand*{\extend}[3]{#1[#2 \mapsto #3]}

\newcommand*{\bang}{\mathord{!}}
\newcommand*{\derives}[2]{#1 \mathrel{::} #2}
\newcommand*{\derivesShade}[2]{\lowlight{#1} \mathrel{\lowlight{::}} #2}

\newcommand*{\slice}[1]{%
\setlength{\fboxsep}{2pt}%
\hspace*{-\fboxsep}\colorbox{lightgray}{\textcolor{gray}{#1}}\hspace*{-\fboxsep}%
}

\newcommand*{\tapp}[2]{#1\,#2}
\newcommand*{\tappBody}[5]{{#1\,#2}\triangleright{#3(#4).#5}}
\newcommand*{\tassign}[2]{#1 := #2}
\newcommand*{\tassignl}[3]{#1 :=_{#2} #3}
\newcommand*{\tcase}[3]{\mathtt{case}\;{#1}\;\mathtt{of}\;\{#2;#3\}}
\newcommand*{\tcaseInl}[4]{\mathtt{case_L}({#1},{#2}.{#3},{#4})}
\newcommand*{\tcaseInr}[4]{\mathtt{case_R}({#1},{#2},{#3}.{#4})}
\newcommand*{\tclosure}[4]{\langle #1, \tfun{#2}{#3}{#4}\rangle}
\newcommand*{\tderef}[1]{\bang #1}
\newcommand*{\tderefl}[2]{\bang_{#1} #2}

\newcommand*{\tfun}[3]{\mathtt{rec}\;#1(#2).#3}
\newcommand*{\tfst}[1]{\mathtt{fst}\;{#1}}
\newcommand*{\thole}[2]{\Hole_{#1}^{#2}}

\newcommand*{\tif}[3]{\mathtt{if}\;{#1}\;\mathtt{then}\;{#2}\;\mathtt{else}\;{#3}}

\newcommand*{\tinl}[1]{\mathtt{inl}\;#1}
\newcommand*{\tinlC}[2]{\mathtt{inl}\;#1.#2}
\newcommand*{\tinr}[1]{\mathtt{inr}\;#1}
\newcommand*{\tinrC}[2]{\mathtt{inr}\;#1.#2}
\newcommand*{\tletFail}[1]{\mathtt{let_F}({#1})}
\newcommand*{\tletSucc}[3]{\mathtt{let_S}({#2}, {#1}.{#3})}
\newcommand*{\tlet}[3]{\mathtt{let}\;{#1=#2}\;\mathtt{in}\;{#3}}
\newcommand*{\twhile}[2]{\mathtt{while}\;{#1}\;\mathtt{do}\;{#2}}
\newcommand*{\ttry}[3]{\mathtt{try}~{#1}~\mathtt{with}\;{#2}\to {#3}}
\newcommand*{\ttryFail}[3]{\mathtt{try_F}({#1},{#2}.{#3})}
\newcommand*{\ttrySucc}[1]{\mathtt{try_S}({#1})}
\newcommand*{\tpair}[2]{{\langle {#1},{#2} \rangle}}
\newcommand*{\traise}[1]{\mathtt{raise}\;{#1}}
\newcommand*{\tref}[1]{\mathtt{ref}\;{#1}}
\newcommand*{\trefl}[2]{\mathtt{ref}_{#1}\;{#2}}
\newcommand*{\tret}[1]{\mathtt{return}\;{#1}}
\newcommand*{\tsnd}[1]{\mathtt{snd}\;{#1}}

\newcommand*{\tunit}{()}
\newcommand*{\varr}[2]{{#1}\{#2\}}

\newcommand*{\tarrayl}[3]{\mathtt{array}_{#1}(#2,#3)}
\newcommand*{\tarrgetl}[3]{#1[#2]_{#3}}
\newcommand*{\tarrsetl}[4]{#1[#2] \gets_{#3} #4}
\newcommand*{\tarray}[2]{\mathtt{array}(#1,#2)}
\newcommand*{\tarrget}[2]{#1[#2]}
\newcommand*{\tarrset}[3]{#1[#2] \gets #3}

\newcommand*{\rret}{\mathtt{val}~}
\newcommand*{\rraise}{\mathtt{exn}~}

\newcommand*{\eval}{\mathrel{\Rightarrow}}
\newcommand*{\writes}{\mathsf{writes}}
\newcommand*{\res}{\mathsf{outcome}}

\renewcommand*{\ell}{\mathcal{l}} 
\newcommand{\kay}{\mathcal{k}}
\newcommand*{\Ell}{\mathcal{L}}

\newcommand*{\bwd}{\searrow}
\newcommand*{\fwd}{\nearrow}
\newcommand*{\bwdFs}[1]{\mathsf{bwd}_{#1}}
\newcommand*{\fwdFs}[1]{\mathsf{fwd}_{#1}}
\newcommand*{\bwdF}{\mathsf{bwd}}
\newcommand*{\fwdF}{\mathsf{fwd}}

\newcommand*{\join}{\sqcup}
\newcommand*{\meet}{\sqcap}
\newcommand*{\sqleq}{\sqsubseteq}
\newcommand*{\sqgeq}{\sqsupseteq}
\newcommand*{\compatible}{\uparrow} 

\DeclareMathOperator{\dom}{dom}

\newcommand*{\antirestrict}{\triangleleft}
\newcommand*{\set}[1]{\{#1\}}

\newcommand*{\update}[3]{#1[#2\mapsto#3]}
\newcommand*{\seqEmpty}{\varepsilon}
\newcommand*{\lowlight}[1]{\textcolor{gray}{#1}}

\newcommand*{\Hole}{\Box}
\newcommand*{\param}{\cdot}

\newcommand*{\appref}[1]{%
\ifdefined\includeapp%
Appendix~\ref{app:#1}%
\else%
an appendix in the full version of the paper%
\fi%
}

\newcommand*{\corref}[1]{Corollary~\ref{cor:#1}}

\newcommand*{\eqnref}[1]{Equation~\ref{eqn:#1}}
\newcommand*{\figref}[1]{Figure~\ref{fig:#1}}
\newcommand*{\lemref}[1]{Lemma~\ref{lem:#1}}
\newcommand*{\thmref}[1]{Theorem~\ref{thm:#1}}

\newcommand*{\Prefix}[1]{{\downarrow}(#1)}

\newenvironment{smathpar}{\begin{mathpar}\small}{\end{mathpar}}
\newenvironment{ruleblock}[1]{%
\begin{flushleft}\fbox{$#1$}\end{flushleft}\begin{smathpar}}{\end{smathpar}%
}

\newcommand*{\caseLocalRaw}{\text{\ding{118}}}
\newcommand*{\caseLocal}{\caseLocalRaw\;}
\newcommand*{\qedLocalRaw}{\text{\ding{234}}}
\newcommand*{\qedLocal}{\qedLocalRaw\;}
\newcommand*{\caseDerivation}[2]{%
\textbf{\emph{Case}}\ %
\begin{adjustbox}{varwidth=#1}%
#2
\end{adjustbox}%
}
\newcommand*{\derivation}[2]{%
\begin{adjustbox}{varwidth=#1}%
#2
\end{adjustbox}%
}

\newcommand*{\proofContext}[1]{\def\currentprefix{proof:#1}}
\newcommand*{\locallabel}[1]{\label{\currentprefix:#1}}
\newcommand*{\localref}[1]{\ref{\currentprefix:#1}}

\newcommand{\crossrule}{\noindent\textcolor{lightgray}{\cleaders\hbox{.}\hfill}}

\newcommand{\BT}{{\bar{T}}}

\newcommand{\rulename}[1]{(\textsc{#1})}

\let\cite=\citep
\bibpunct{[}{]}{;}{a}{}{,}

\begin{document}

\acmJournal{PACMPL}

\title{Imperative Functional Programs that Explain their Work}

\author{Wilmer Ricciotti}
\affiliation{
  \department{Laboratory for Foundations of Computer Science}              
  \institution{University of Edinburgh}            
}
\email{wilmer.ricciotti@ed.ac.uk} 

\author{Jan Stolarek}
\affiliation{
 \department{Laboratory for Foundations of Computer Science}              
  \institution{University of Edinburgh}            
}
\email{jan.stolarek@ed.ac.uk} 

\author{Roly Perera}
\affiliation{
 \department{Laboratory for Foundations of Computer Science}              
  \institution{University of Edinburgh}            
}
\email{roly.perera@ed.ac.uk} 

\author{James Cheney}
\affiliation{
  \department{Laboratory for Foundations of Computer Science}              
  \institution{University of Edinburgh}            
}
\email{jcheney@inf.ed.ac.uk} 

\begin{abstract}
  Program slicing provides explanations that illustrate how program
  outputs were produced from inputs.  We build on an
  approach introduced in prior work by \citet{perera12icfp}, where dynamic slicing was defined
  for pure higher-order functional programs as a Galois connection
  between lattices of partial inputs and partial outputs.  We extend
  this approach to \emph{imperative functional programs} that combine
  higher-order programming with references and exceptions.  We present
  proofs of correctness and optimality of our approach and a
  proof-of-concept implementation and experimental evaluation.
\end{abstract}



\maketitle

\section{Introduction}
\label{sec:intro}

When a program produces an unexpected result, experienced programmers often
intuitively ``run the program backwards'' to identify the last part of the
program that contributed to the output, to decide where to focus
attention to find the bug.  Effects such as mutable state (references, arrays)
or exceptions can make this a nontrivial challenge.  For example, in the following ML-like program
(where \verb|!y| means the value of reference cell \verb|y|):
\begin{verbatim}
   let f(x) = if (x == 0) then y := 6 * !z else (y := 84 / !z; w := g(!y + 12))
\end{verbatim}
suppose we observe, on applying \verb|f| to some argument, that afterwards
\verb|!y| has the value \verb|42|, when we were expecting some other value.  We
might reasonably focus on the two possible assignments in the
$\mathtt{else}$ branch, and
hypothesise that perhaps the strange output resulted from \verb|x| having the
value \verb|1| and reference cell \verb|z| containing \verb|2|. Of course,
this reasoning relies on certain working assumptions, which may be invalid: we
do not know whether \verb|w| and \verb|y| are aliases, we do not know whether
\verb|g| had side-effects, and we do not know the value of \verb|x| that
determined which branch was taken. Furthermore, if the above code executed
inside an exception handler:
\begin{verbatim}
   try f(1) with Division_by_zero -> y := 42
\end{verbatim}
then there is another possible explanation: perhaps \verb|!z| is \verb|0|, so
the attempt to divide \verb|84| by zero failed, raising an exception whose
handler eventually assigned \verb|42| to \verb|y|.  Alternatively, such an
exception could have been raised from within the function \verb|g|.  This
illustrates that the exact sequence of events leading to an unexpected result
may be impossible to determine based solely on the output. Often, programmers
resort to interactive debugging, or even manually adding \verb|print| statements,
to observe the actual control flow path.

Backwards reasoning to identify the parts of a program that may have
contributed to the output is the basis of a popular analysis technique called
\emph{program slicing}, invented by \citet{weiser81icse}.  In general, slicing
techniques take a program and a \emph{slicing criterion} describing the part
of the program's output or behaviour of interest, and produce a \emph{slice},
or subset of the program text, that identifies the parts of the program
relevant to the criterion, omitting (as much as possible) parts that were not
relevant. A typical slicing criterion might consist of a source location $P$
(expression or statement) and a set of variables, and the slice would contain
those parts of the program deemed to be relevant to the values of those
variables at $P$.
  
Slicing techniques can be divided into two broad
categories: \emph{static} slicing conservatively analyses all possible
executions of the program and identifies those parts which \emph{potentially}
influence the slicing criterion, whereas \emph{dynamic} slicing identifies
those parts of the program which \emph{do} influence the slicing criterion on
a particular execution. Because dynamic slicing analyses a specific run, it is
especially useful for debugging and testing, where the goal is to understand a
particular scenario or test case as precisely as possible.

In the above example, a static slicing algorithm might deem it unsafe to
exclude any part of the program from the static slice because any part of the
program could have affected the final value of
\verb|y|. Using dynamic slicing, we can increase precision by considering exactly
what happened, given the actual store that the program ran in. For example, if
we shade elided parts of the program, one possible slice with respect to
output observation
\verb|!y| $=$ \verb|42| is:
\begin{Verbatim}[commandchars=\\\{\}]
   let f(x) = if (x == 0) then \(\slice{y := 6 * !z}\) else (y := 84 / !z; \(\slice{w := g(!y + 12)}\))
   try f(1) with Division_by_zero -> y := 42
\end{Verbatim}
This slice shows that the \verb|Division_by_zero| exception was raised and
handled (so \verb|!z| must have been \verb|0|).  On the other hand, if slicing
yields:
\begin{Verbatim}[commandchars=\\\{\}]
   let f(x) = if (x == 0) then \(\slice{y := 6 * !z}\) else (y := 84 / !z; \(\slice{w}\) := g(!y + 12))
   try f(1) with Division_by_zero -> \(\slice{y := 42}\)
\end{Verbatim}
then this means no exception was raised (so \verb|!z| was not \verb|0|),
\verb|w| did not alias \verb|y|, and the final assignment of \verb|42| to
\verb|y| must have been a side-effect of \verb|g| which used its prior value
of
\verb|84 / !z|.

The slicing used in these examples is \emph{backward} slicing, which works
back from outputs to contributing program parts. \citet{bergeretti85}
introduced the complementary technique of \emph{forward} slicing, which works
forward from program parts to outputs they contribute to. Forward slicing
corresponds to the kind of informal reasoning programmers use during debugging
when they try to understand the \emph{consequences} of a fragment of code.
Needless to say, program slicing, both forward and backward, has turned out to
have many applications in program transformation and optimisation besides
debugging, and has been researched very thoroughly in the context of
mainstream imperative and object-oriented programming languages such as C/C++
and Java; \citet{xu05sigsoft} cite over 500 papers on slicing. Slicing for
functional programs, however, has received comparatively little attention.
\citet{biswas97phd} developed slicing techniques for a higher-order ML-like
language, including references and exceptions, but only with respect to the
whole program result as slicing criterion.  Other authors have investigated
slicing for pure or lazy languages such as Haskell or
Curry~\citep{silva06ppdp,ochoa08hosc,rodrigues07jucs}.

\citet{perera12icfp} introduced a new approach to dynamic slicing for (pure)
functional programs where the slicing criteria take the form of \emph{partial
values}, allowing for fine-grained slicing particular to specific sub-values.
In this approach, input and output values may be partly elided in the same way
as program slices, with \emph{holes} (written $\Hole$) intuitively
corresponding to parts of programs or values that are not of interest.  Perera
\etal showed how to extend the usual semantics of pure, call-by-value programs
with rules for $\Hole$ to construct a \emph{Galois connection} for program
slicing. The forward component of the Galois connection maps a partial input
$x$ to the \emph{greatest} partial output $y$ that can be computed from $x$;
the backward component of the Galois connection maps a partial output $y$ to
the \emph{least} partial input $x$ from which we can compute $y$.  (Note that
this use of Galois connections for \emph{dynamic} slicing is unrelated to
their widespread application to \emph{static} analysis techniques such as
abstract interpretation~\citep{cousot77popl,darais16icfp} or gradual
typing~\citep{garcia16popl}.)  Once the forward slicing semantics is defined,
the behaviour of backward slicing is uniquely determined as the \emph{lower
adjoint} of the forward semantics, whose existence is established using
standard lattice-theoretic techniques.  Perera \etal also showed how to
compute such slices efficiently, through a semantics instrumented with
\emph{traces} that record details of program execution.

In this paper, we build on this fine-grained approach and address the
challenges of adapting it to \emph{imperative functional programming}: that
is, programming with higher-order functions, references, and exceptions. We
focus on a simplified, ML-like core language, so our approach is immediately
relevant to languages such as Standard ML, OCaml, Scheme, or F\#.  Types do
not feature in our approach, so our results should apply to both statically
typed and dynamically typed languages. However, this paper focuses on
foundational aspects of slicing for imperative functional programming, via a
core language and proof of correctness, and more work would need to be done to
develop a full-scale slicing tool for a mainstream language.

To illustrate how our approach compares to previous work, here is a
(contrived) program that our slicing system (and no prior work)
handles, and its slice explaining why an exception was raised:

\begin{tabular}{p{6cm}p{6cm}}
\begin{Verbatim}
  let a = ref 1 in
  let b = ref 2 in
  map (fun c -> b := !b - 1 ; 1/!c) 
      [a,b]
\end{Verbatim}
&
\begin{Verbatim}[commandchars=\\\{\}]
  let a = \(\slice{ref 1}\) in
  let b = ref 2 in
  map (fun c -> b := !b - 1 ; 1/!c) 
      [\(\slice{a}\),b]
\end{Verbatim}
\end{tabular}

This program does not return normally; it raises an exception because of the
attempt to divide \verb|1| by zero.  Our approach produces a
backward slice (shown on the right) as an explanation of the exception.
It shows that the exception was raised because of the attempt to
divide \verb|1| by \verb|!c|, when \verb|!c| was zero after \verb|b|
was decremented the second time. In Biswas' approach (the only prior
work to handle higher-order functions, references, and exceptions),
the whole program would have to be included in the slice: without the
ability to represent slicing criteria as partial values, there is no
way to capture the partial usage of the list value supplied to
\verb|map| (in other words, the fact that only part of the list was
needed to produce the exception). Our approach, in contrast, allows us
to slice each sub-computation with respect to a precise criterion
reflecting exactly the contribution required for that step. Here, it
is safe to slice away the expression that defined \verb|a| as long as
we remember that it did not throw an exception.  The main contribution
of this paper is showing how to make the above intuitions precise and
extend the Galois connection approach to higher-order programming with
effects.



\subsection{Contributions and Outline}

In the rest of this paper, we present the technical details of our
approach together with a proof-of-concept implementation.
In detail, our contributions are as follows:
\begin{itemize}
\item (Section~\ref{sec:problem}) We first review (what we call)
  \emph{Galois slicing}, the Galois connection approach to dynamic slicing
  introduced by \citet{perera12icfp}, illustrated using a simple expression
  language.  In particular, we offer a direct argument showing that any pair
  of forward and backward slicing functions that satisfy appropriate
  optimality properties form a Galois connection
  (Prop.~\ref{prop:optimal-slicing-galois}).
\item (Section~\ref{sec:calculus}) We extend Perera \etal's core language TML
  (Transparent ML) with exceptions and references, and call the result iTML
  (``imperative TML'').  Our core language differentiates between pure
  \emph{expressions} and \emph{computations} that may have side effects.  We
  define partial values, expressions, and traces, and state the rules
  (similar to those of Perera \etal) for slicing pure expressions.
\item (Section~\ref{sec:coarse-slicing}) We define forward and
  backward slicing semantics for effectful computations, which abstract away
  information about results of sub-computations that return normally and do
  not affect the slicing criterion. The Galois connection for a computation
  includes the proof term, or \emph{trace}, that explains how the result was
  computed, allowing the slicing semantics to compute a \emph{least
  explanation} optimised to a particular partial outcome.
Thm.~\ref{thm:galois-connection:computation} proves this and is our
main technical contribution.
\item (Section~\ref{sec:extensions}) We consider several natural
  extensions: slicing in the presence of mutable arrays, while-loops
  and sequential composition, and give illustrative examples.
\item (Section~\ref{sec:implementation}) We present a proof-of-concept
  implementation of our approach in Haskell, discuss a more
  substantial example,  and make preliminary observations about performance.
\end{itemize}
In the remainder of the paper, we discuss related work and future
directions in greater detail, and summarise our findings.  Detailed
proofs of our main results, some straightforward rules, and an
extended example are all included in the %
\ifdefined\includeapp%
appendices.%
\else%
appendices of the full version of the paper.%
\fi

\section{Background: Galois slicing}
\label{sec:problem}

In this section we recapitulate the Galois connection approach to dynamic
slicing introduced for pure functional programs by \citet{perera12icfp}, using
a simple expression language as an example. We call their approach
\emph{Galois slicing}. We then discuss the challenges to adapting this
framework to references and exceptions; the rest of the paper is a concrete
instantiation of this framework in that setting.

\subsection{Ordered sets, lattices and Galois connections}

We first review ordered sets, lattices and Galois connections.  An
\emph{ordered set} $(P,\leq)$ is a set $P$ equipped with a partial order
$\leq$, that is, a relation which is reflexive, transitive and antisymmetric.
A function $f : P \to Q$ between ordered sets $(P,\leq_P)$ and $(Q,\leq_Q)$ is
\emph{monotone} if it preserves the partial order, i.e.~if $x \leq_P x'$
implies $f(x) \leq_Q f(x')$ for all $x,x' \in P$.  The \emph{greatest lower
bound} (or \emph{meet}) of two elements $x,y\in P$ (if it exists) is written
$x \meet y$ and is the largest element of $P$ such that $x
\geq x \meet y \leq y$.  The \emph{least upper bound} (or
\emph{join}) $x \join y$ is defined dually as the least element satisfying $x
\leq x \join y \geq y$ when it exists.  Likewise we write $\bigsqcup S$ or
$\bigsqcap S$ for the least upper bound or greatest lower bound of a subset
$S$ of $P$, when it exists.

A \emph{lattice} is an ordered set in which all pairwise meets and joins
exist. A lattice is \emph{complete} if all subsets have a meet and join, and
\emph{bounded} if it has a least element $\bot$ and a greatest element $\top$.
All finite lattices are complete and bounded, with $\bot = \bigsqcap P$ and
greatest element $\top = \bigsqcup P$. A function $f : P \to Q$ where $Q$ is a
lattice is \emph{finitely supported} if $\{x\in P \mid f(x) \neq \bot\}$ is
finite, where $\bot$ is the least element of $Q$. Given $x \in P$, we define
the \emph{lower set} $\Prefix{x} = \{x' \in P \mid x'\leq x\}$ of all elements
below $x$.  We introduce the notion of a \emph{partonomy} for a set $X$,
which we define to be a partial order $P_X \supseteq X$ such that every
element of $X$ is maximal in $P_X$ and $\Prefix{x}$ is a finite lattice for
all elements $x \in X$.

Given ordered sets $(P,\leq_P)$ and $(Q,\leq_Q)$, a \emph{Galois connection}
is a pair of (necessarily monotone) functions $(f : P \to Q,g : Q \to P)$ that
satisfy
\[f(p) \leq_Q q \iff p \leq_P g(q)\]
The function $f$ is sometimes called the \emph{lower adjoint} and $g$ the
\emph{upper adjoint}.  We say $f$ and $g$ are \emph{adjoint} (written $f
\dashv g$) when $(f,g)$ is a Galois connection with lower adjoint $f$ and
upper adjoint $g$.

\subsection{Galois connections for slicing}

Now we show how to interpret \emph{partial programs} and \emph{partial values}
in this setting of lattices and Galois connections. We consider a simple
language of expressions with numbers, addition, and pairs.
\[\begin{array}{rclcrcl}
e &::=& n \mid e_1 + e_2 \mid (e_1,e_2) \mid \tfst{e} \mid \tsnd{e} &&
v &::=& n \mid (v_1,v_2)
\end{array}\]
Suppose a standard big-step semantics is defined using a judgment $e
\eval v$, as follows:
\begin{smathpar}
\inferrule*{\strut}
{n \eval n}
\and
\inferrule*{e_1 \eval n_1\\
e_2 \eval n_2}{
e_1+ e_2 \eval n_1 +_{\mathbb{N}} n_2}
\and
\inferrule*{e_1 \eval v_1\\
e_2 \eval v_2}{
(e_1, e_2) \eval (v_1,v_2)}
\and
\inferrule*{e \eval (v_1,v_2)}{
\tfst{e} \eval v_1}
\and
\inferrule*{e \eval (v_1,v_2)}{
\tsnd{e} \eval v_2}
\end{smathpar}

\vspace{-8pt} 
\noindent A \emph{partial} expression is an expression that may contain a
\emph{hole} $\Hole$. We define an order on partial expressions as follows.
(From now on we write the partial orders simply as $\sqleq$, omitting the
subscripts.)
\begin{smathpar}
\inferrule*{\strut}{\Hole \sqleq e}
\and
\inferrule*{\strut}{n \sqleq n}
\and
\inferrule*{e_1 \sqleq e_1' \\
e_2 \sqleq e_2'}{
e_1 + e_2 \sqleq e_1' + e_2'}
\and
\inferrule*{e_1 \sqleq e_1' \\
e_2 \sqleq e_2'}{
(e_1, e_2) \sqleq (e_1', e_2')}
\and
\inferrule*{e \sqleq e'}{
\tfst{e} \sqleq \tfst{e'}}
\and
\inferrule*{e \sqleq e'}{
\tsnd{e} \sqleq \tsnd{e'}}
\end{smathpar}

\vspace{-8pt} 
\noindent
This relation is simply the compatible partial order generated by $\Hole
\sqleq e$, and thus is reflexive, transitive and antisymmetric.  It is easy to
verify that meets $e_1\meet e_2$ exist for any two partial expressions; for
example, $(1,2)
\meet (1,2+2) = (1, \Hole)$; however, joins do not always exist, since for
example there is no expression $e$ satisfying $(1,2) \sqleq e \sqgeq (1,2+2)$.
Nevertheless, if we restrict attention to the (finite) set $\Prefix{e}$ of
prefixes of a given expression $e$, joins do exist; that is, $\Prefix{e}$ is a
(finite) lattice, with $\bot = \Hole$ and $\top = e$.  For example,
$\Prefix{(1,2)} =
\{\Hole,(1,\Hole),(\Hole,2),(1,2)\}$, and the join $(1, \Hole) \join (\Hole,
2)$ is defined and equal to $(1, 2)$.  Since all values happen to be
expressions, we can also derive lattices $\Prefix{v}$ of prefixes of a given
value $v$, obtaining $\sqleq$ by restriction from the partial order on
expressions.

Suppose we have some ``computation'' relation $C\subseteq X \times Y$, which
for now we assume to be deterministic.  Given a particular computation $(x,y)
\in C$, we would like to define a technique for slicing that computation.  We
start by defining partonomies $P_X$ and $P_Y$ for $X$ and $Y$.  Given a
partial input $x' \in \Prefix{x}$, it is natural to expect there to be a
corresponding partial output $y' \in \Prefix{y}$ that shows how much of the
output $y$ can be computed from the information available in $x'$. Suppose
such a function $\fwdF
: \Prefix{x} \to \Prefix{y}$ is given. We already know that given \emph{all}
of the input ($x \in \Prefix{x}$) we can compute all of the output $y$. Thus
if $\fwdF$ computes as much as possible, then $\fwdF(x)$ should certainly be
$y$. By the same token, it seems reasonable that given \emph{none} of the
input ($\Hole
\in
\Prefix{y}$) we should be unable to compute any of the output, so that
$\fwdF(\Hole) =
\Hole$. More generally, we would like to be able to compute \emph{partial}
output from \emph{partial} input, so that for example $\fwdF(\Hole,2+2) =
(\Hole,4)$ since we can compute the second component of a pair without any
knowledge of the first. Finally, a reasonable intuition seems to be that
$\fwdF$ should be monotone, since learning more
information about the input should not make the output any less certain.

As a concrete example of such a $\fwdF$ function, we can extend the
(deterministic) evaluation relation $\eval$ defined earlier for expressions to
\emph{partial} expressions as follows:
\begin{smathpar}
\inferrule*{\strut}
{n \fwd n}
\and
\inferrule*{e_1 \fwd n_1\\
e_2 \fwd n_2}{
e_1+ e_2 \fwd n_1 + n_2}
\and
\inferrule*{e_1 \fwd v_1\\
e_2 \fwd v_2}{
(e_1, e_2) \fwd (v_1,v_2)}
\and
\inferrule*{e \fwd (v_1,v_2)}{
\tfst{e} \fwd v_1}
\and
\inferrule*{e \fwd (v_1,v_2)}{
\tsnd{e} \fwd v_2}
\\
\inferrule*{\strut}
{\Hole \fwd \Hole}
\and
\inferrule*{e_1 \fwd \Hole}
{e_1+e_2 \fwd \Hole}
\and
\inferrule*{e_1 \fwd v_1 \\
e_2 \fwd \Hole}
{e_1+e_2 \fwd \Hole}
\and
\inferrule*{e \fwd \Hole}{
\tfst{e} \fwd \Hole}
\and
\inferrule*{e \fwd \Hole}{
\tsnd{e} \fwd \Hole}
\end{smathpar}

\vspace{-8pt} 
\noindent 
If the parts needed to perform a given evaluation step are present, we behave
according to the corresponding $\eval$ rule (top row). Otherwise we use a rule
from the bottom row: if the expression itself is missing ($\Hole$), or if a
needed intermediate result (e.g.~the tuple value for a projection, or an
argument to an addition) is $\Hole$, then the result is again $\Hole$. So, for
example, $(\Hole + 1, 1) \fwd (\Hole,1)$.  For a given $e \eval v$ and $e' \in
\Prefix{e}$, it is not possible for $e' \fwd$ to get stuck, for example by
trying to add an integer to a pair, and indeed one can verify that these rules
define a total, monotone function $\fwdFs{e}: \Prefix{e} \to \Prefix{v}$ such
that $\fwdFs{e}(e') = v' \iff e' \fwd v'$. The function $\fwdFs{e}$ computes
the \emph{forward slice} of $e' \in \Prefix{e}$, namely that portion of $v$
which can be computed using only the information in $e'$.

Now, given $\fwdFs{e}: \Prefix{x} \to \Prefix{y}$, we would like to define a
converse mapping $\bwdFs{e}: \Prefix{y} \to \Prefix{x}$ which computes the
\emph{backward slice} for $y' \in
\Prefix{y}$, namely a partial input large enough to compute $y'$. That is, we
want $\bwdFs{e}$ to satisfy $y' \sqleq \fwdFs{e}(\bwdFs{e}(y'))$ for any $y'
\in \Prefix{y}$.  We call this property \emph{consistency}.

There may be many consistent choices of $\bwdFs{e}$.  As an extreme example,
the constant function $\bwdF_0(y') = x$, which simply produces the full input
for any output prefix, is consistent.  However, as a slicing function it is
singularly useless: it treats all parts of the input as relevant, failing to
take advantage of the fact that not all of the output was required. Ideally,
therefore, we would like $\bwdF$ to satisfy the following
\emph{minimality} property:
\begin{equation}
\label{eqn:minimality}
\bwdFs{e}(y') = \bigsqcap \{x' \mid y' \sqleq \fwdFs{e}(x')\}
\end{equation}
This (together with consistency) says that $\bwdFs{e}(y')$ is the smallest part of
the input that provides enough information to recompute $y'$ using $\fwdFs{e}$.

Now, if $\bwdFs{e}$ is the lower adjoint of $\fwdFs{e}$ (if $\bwdFs{e}$ and
$\fwdFs{e}$ form a Galois connection $\bwdFs{e} \dashv
\fwdFs{e}$) then the monotonicity, consistency and minimality properties
follow by standard arguments~\citep{davey}.  More surprisingly, these
properties suffice to ensure that $\bwdFs{e} \dashv \fwdFs{e}$:

\begin{proposition}\label{prop:optimal-slicing-galois}
  Given complete lattices $P,Q$, suppose $g : Q \to P$ is monotone and $f : P
  \to Q$ is consistent and minimal with respect to $g$.  Then they form a
  Galois connection $f \dashv g$.
\end{proposition}
\begin{proof}
  First suppose $f(p) \sqleq q$.  Then $p \sqleq g(f(p))
  \sqleq g(q)$ by consistency of $f$ and monotonicity of $g$.  This proves
  that $f(p) \sqleq q \Rightarrow p \sqleq g(q)$. For the converse, assume
  that $p \sqleq g(q)$.  Then
\[f(p) = \bigsqcap \{q' \mid p \sqleq g(q')\} \sqleq q\]
where the equality is the minimality of $f$ and the inequality holds because
$p \sqleq g(q)$.  This proves that $p \sqleq g(q) \Rightarrow f(p) \sqleq q$,
so $f \dashv g$.
\end{proof}

It may appear difficult to design an adjoint pair of functions $\fwdFs{e}$ and
$\bwdFs{e}$, or even to be sure that one exists for a given candidate
definition of $\fwdFs{e}$.  Luckily, another standard result applies: if $P$
is a complete lattice then $g : Q\to P$ has a lower adjoint $f
: P \to Q$ if and only if $g$ preserves meets, that is, $g(\bigsqcap S) =
\bigsqcap \set{g(s) \mid s \in S}$.  For finite lattices, it suffices to
consider only binary meets and the top element: $g(q_1 \meet q_2) =
g(q_1) \meet g(q_2)$ and $g(\top_Q) = \top_P$. Moreover, the lower adjoint $f
: P \to Q$ is uniquely determined by the minimality equation. (Dually, any
\emph{join}-preserving function between complete lattices uniquely determines
a ``maximising'' meet-preserving function as its upper adjoint, but for an
evaluation relation it seems more natural to start with forward slicing and
induce the backward-slicing function.)

Of course, for a given computation there may be more than one choice of
lattice structure for the input and output, and there may also be more than
one natural choice of meet-preserving forward slicing.  Once such choices are
made, however, a minimising backward-slicing function is determined by the
forward semantics. Nevertheless, there are two considerations
(beyond meet-preservation) that make defining a suitable forward-function
non-trivial: first, the availability of an efficient technique for computing
the backward-slicing lower adjoint, and second, the precision of the
forward-slicing function (which in turn determines the precision of backward
slicing).

To see the first point, we return to our simple expression language. It is
easily verified that $\fwdFs{e}$ is indeed meet-preserving in that
$\fwdFs{e}(e_1
\meet e_2) =
\fwdFs{e}(e_1) \meet \fwdFs{e}(e_2)$ and $\fwdFs{e}(e) = v$. Since
$\Prefix{e}$ is a finite lattice, it follows that $\fwdFs{e}$ has a lower
adjoint $\bwdFs{e} : \Prefix{v} \to
\Prefix{e}$, which satisfies the minimality property (\eqnref{minimality}).
The minimality property alone, however, is not suggestive of an efficient
procedure for computing $\bwdFs{e}$.  Read naively, it suggests evaluating
$\fwdFs{e}$ on all partial inputs (of which there may be exponentially many in
the size of $e$), and then computing the meet of all partial inputs $e'$
satisfying $v' \sqleq \fwdF(e')$. 

Perera \etal showed how the lower adjoint can be efficiently computed by
adopting an algorithmic style ``dual'' to forward slicing: whereas forward
slicing \emph{pushes $\Hole$ forward} through the computation, erasing any outputs
that depend on erased inputs, backward slicing can be implemented by \emph{pulling
$\Hole$ back} through the computation, erasing any parts of the input
which are not needed to compute the required part of the output. In their
functional setting, this involves using a \emph{trace} of the computation to
allow the slicing to proceed backwards. In the toy language we consider here,
the expression itself contains enough information to implement backward
slicing in this style. The following definition illustrates:
\[\begin{array}{rclcrcl}
\bwdFs{n}(n) &=& n
&&
\bwdFs{e_1 + e_2}(n) & = & \bwdFs{e_1}(\fwdF(e_1)) + \bwdFs{e_2}(\fwdF(e_2))
\\
\bwdFs{\tfst{e}}(v) &=& \bwdFs{e}(v,\Hole)
&&
\bwdFs{(e_1,e_2)}(v_1,v_2) &=& (\bwdFs{e_1}(v_1), \bwdFs{e_2}(v_2)) 
\\
\bwdFs{\tsnd{e}}(v) &=& \bwdFs{e}(\Hole,v)
&&
\bwdFs{e}(\Hole) &=& \Hole
\end{array}\]
The interesting cases are those for addition and projections.  For addition,
we continue slicing backwards, using the original values of the
subexpressions.  This still leaves something to be desired, since we use
$\fwdF$ to reevaluate subexpressions $e_1$ and $e_2$. (It is possible to avoid
this recomputation in the trace-based approach by recording extra information
about the forward evaluation that $\bwdF$ can use.) For projections such as
$\tfst{e}$, we slice the subexpression $e$ with respect to partial value
$(v,\Hole)$, expressing the fact that we do not need the second component.  If
the partial output is a hole, then we do not need any of the input to
recompute the output.  For example, suppose $e = (1,\tfst{(1,2)}+3)$. Then
$\bwdFs{e}(\Hole,4)$ yields $(\Hole,
\tfst(1,\Hole)+3)$ because we do not need the first $1$ or the $2$ to
recompute the result. One of the key challenges we address in
this paper is adapting this algorithmic style to deal with imperative features
like exceptions and stores.

To see the importance of precision for forward slicing, consider the following
alternative slicing rules for pairs:
\begin{smathpar}
\inferrule*
{
  e_1 \fwd \Hole
}
{
  (e_1, e_2) \fwd (\Hole,\Hole)
}
\and
\inferrule*[right={$v_1 \neq \Hole$}]
{
  e_1 \fwd v_1
  \\
  e_2 \fwd v_2
}
{
  (e_1, e_2) \fwd (v_1,v_2)
}
\end{smathpar}

\vspace{-8pt} 
\noindent Evaluation goes left-to-right, and so naively we might suppose that
if we know nothing about the first component, we should not proceed with the
second component. The forward-slicing function for a given computation still
preserves meets, and thus has a backward-slicing lower adjoint. However, again
supposing $e = (1,\tfst{(1,2)}+3)$, we have $\fwdFs{e}(\Hole,\tfst{(1,2)}+3) =
(\Hole,\Hole)$ by the first rule above. If we then use this partial output to
backward slice, we find $\bwdFs{e}(\Hole,\Hole) = (\Hole,\Hole)$: if all we
need of the output is the fact that it is a pair, then all we actually needed
of the program was the fact that it computes a pair. Thus the ``round trip''
$\bwdFs{e}(\fwdFs{e}(e'))$, technically a \emph{kernel operator}, reveals
\emph{all} the parts of $e$ that are rendered irrelevant as a consequence of
retaining only the information in $e'$, and here $\bwdFs{e}(\fwdFs{e}(e'))$
reveals a (spurious) dependency of the second component of the pair on the
first. This motivates the more precise pair-slicing rule we first presented,
which sliced each component independently. A key design criterion for
forward-slicing therefore is that it only reflect genuine dependencies,
capturing specifically how input was consumed in order to produce output.
Defining suitably precise forward slicing in the presence of stores and
exceptions is another of the key challenges we address in this
paper.

\subsection{Summary}

To summarise, the Galois slicing framework involves the following steps:
\begin{itemize}
\item Given sets of expressions, values and other syntactic objects,
  define  partonomies such that the set of prefixes of each object
  forms a finite lattice.
\item Given a reference semantics for the language, say a deterministic
  evaluation relation $ {\eval} \subseteq X \times Y$, define a family of
  meet-preserving functions $\fwdFs{x} : \Prefix{x} \to \Prefix{y}$ for every
  $x \eval y$. By the reasoning given above, $\fwdFs{x}$ has a lower adjoint
  $\bwdFs{x}:
  \Prefix{y} \to \Prefix{x}$ that computes the least slice of the input that
  suffices to recompute a given partial output.
\item Define a procedure for computing a backward slice, typically by running
back along a trace of $x \eval y$, and show that the procedure computes the
lower adjoint $\bwdFs{x}$.
\end{itemize}

\noindent The framework describes a design space for optimal slicing
techniques for a given language: we have latitude to decide on
suitable lattices of partial inputs and outputs and a suitable
definition of forward slicing, as long as it is compatible with
ordinary evaluation and is a meet-preserving function.  The definition
of forward slicing must be precise enough to reflect accurately how
information in the input is consumed during execution to produce
output, and there may be different notions of slicing suitable for
different purposes. Once these design choices are made, the
extensional behaviour of an optimal backward slicing $\bwdF$ is
determined, and the remaining challenge is to find an efficient method
for computing backward slices, using traces where appropriate.

\section{Core calculus and common concepts}
\label{sec:calculus}

\begin{figure}[t]
\[\small\begin{array}{lrcll}
\textbf{Expression} 
& e 
& ::= 
& x \mid \tunit \mid \tinl{e} \mid \tinr{e} \mid \tpair{e_1}{e_2} \mid \tfst{e} \mid \tsnd{e} \mid \tfun{f}{x}{M}  &
                                                                   \mid \Hole
\\

\textbf{Computation} 
& M, N 
& ::= 
& \tret{e} \mid  \tlet{x}{M_1}{M_2}\mid  \tapp{e_1}{e_2} \\
&&\mid&\tcase{e}{\tinlC{x}{M_1}}{\tinrC{y}{M_2}}\\
&&\mid& \traise\;e \mid \ttry{M_1}{x}{M_2} \\
&&\mid& \tref{e} \mid \tassign{e_1}{e_2} \mid \bang e  & \mid \Hole
\smallskip\\

\textbf{Environment} 
& \rho, \sigma
& ::=
& \seqEmpty \mid \update{\rho}{x}{v}
\\

\textbf{Store} 
& \mu, \nu
& ::=
& \seqEmpty \mid \update{\mu}{\ell}{v}
\\
\textbf{Set of locations} 
& \Ell
& ::=
& \{\ell_1,\ldots,\ell_n\}
\\

\textbf{Value} 
& u, v 
& ::=
& \tunit \mid \tinl{v} \mid \tinr{v} \mid \tclosure{\rho}{f}{x}{M}
  \mid \ell  
& \mid \Hole
\\

\textbf{Outcome}
& \kay
& ::=
& \rret \mid \rraise
\\

\textbf{Result}
& R
& ::=
& \kay~v 
\smallskip\\

\textbf{Trace}
& T, U
& ::= 
& \tret{e} \mid \tletFail{T} \mid \tletSucc{x}{T_1}{T_2} \mid
  \tappBody{e_1}{e_2}{f}{x}{T} \\ 
&&\mid&  \tcaseInl{e}{x}{T}{y} \mid \tcaseInr{e}{x}{y}{T} \\
&&\mid& \traise{e} \mid\ttrySucc{T} \mid 
\ttryFail{T_1}{x}{T_2} \\
&&\mid& \trefl{\ell}{e} \mid \tassignl{e_1}{\ell}{e_2}  \mid
        \tderefl{\ell}{e}
& \mid \thole{\Ell}{\kay}
\end{array}\]
\caption{Abstract syntax}
\label{fig:syntax}
\end{figure}

We now introduce the core calculus iTML, ``Imperative Transparent ML'', which
extends the TML calculus of \citet{perera12icfp} with ML-like references and
exceptions. These features potentially complicate an operational semantics,
since any subexpression might modify the state or raise an exception.  To
avoid a proliferation of rules and threaded arguments, and to help illuminate
the underlying ideas, we present the language using a variant of fine-grained
call-by-value \citep{levy03}, which distinguishes between (pure)
\emph{expressions} and (effectful) \emph{computations}. The syntax of the
calculus, including runtime constructs, is presented in
Figure~\ref{fig:syntax}. We omit typing rules, since static types currently
play no role in the Galois slicing approach.  Likewise, we omit constructs
associated with isorecursive types since they contribute little in the absence
of a type system.  In our implementation, the source language is typed and we
consider a fixed type for exceptions (for the moment, this is
$\texttt{string}$, but any other type, such as ML's extensible exception type,
would also work).

Usually we can consider a core calculus with separate expressions and
computations without loss of generality because general programs can
be handled by desugaring. However, since our goal is to produce slices
of the original program, slicing desugared programs would necessitate
resugaring slices (that is, translating them back to slices of the
original program, following \citet{pombrio14pldi,pombrio15icfp}).
Rather than pursue this indirect approach, our implementation handles
general programs directly, extrapolating from the core calculus
presented in this paper. We give further details in
Section~\ref{sec:implementation}.
 
The expression forms include variables, introduction forms for unit, pairs,
sums, recursive functions, and side-effect-free elimination forms (pair
projections). Computations represent full programs with effects: expressions
are lifted to computations using the $\mathtt{return}$ operator (corresponding
to monadic return), which represents the program terminating normally and
returning the value of that expression, and computations are composed using
$\tlet{x}{M}{N}$, which corresponds to monadic bind. Abnormal termination is
initiated by $\traise{e}$, which raises $e$ as an exception and aborts the
current computation. Other computation forms include exception handling
$\mathtt{try}/\mathtt{with}$, reference cell creation $\tref{e}$,
dereferencing $\tderef{e}$, assignment $\tassign{e_1}{e_2}$, case analysis,
and function application $\tapp{e_1}{e_2}$.
 
The evaluation of an expression yields a \emph{value}. Values are closed and
include units, pairs, injections, closures $\tclosure{\rho}{f}{x}{N}$, and
locations. The evaluation of a computation yields a \emph{result}, which is
either a success $\rret~{v}$ or failure $\rraise~{v}$, where $v$ is a value.
An \emph{environment} $\rho$ is a finitely supported function from variable
names to values. (Recall that this means that $\rho(x)$ is defined for at most
finitely many $x$.) We write $\update{\rho}{y}{v}$ for the operation which
extends $\rho$ by mapping $y$ to $v$, where $\rho(y)$ was previously
undefined. A \emph{store} is a finitely supported function from locations
$\ell$ to values. Store update $\update{\mu}{\ell}{v}$ is similar to
environment update except that we do not require $\mu(\ell)$ to be undefined,
as the update may overwrite the previous value of $\ell$.

\begin{figure}
\vspace{2mm}
\begin{ruleblock}{\rho, e \eval v}
\inferrule*
{
	x \in \dom(\rho)
}
{
	\rho, x \eval \rho(x)
}
\and
\inferrule*
{
	\strut	
}
{	
	\rho, \tunit \eval \tunit
}
\and
\inferrule*
{
	\strut
}
{
	\rho, \tfun{f}{x}{M} \eval \tclosure{\rho}{f}{x}{M}
}
\and
\inferrule*
{
	\rho, e \eval v
}
{
	\rho, \tinl{e} \eval \tinl{v}
}
\and
\inferrule*
{
	\rho, e \eval v
}
{
	\rho, \tinr{e} \eval \tinr{v}
}
\and
\inferrule*
{
	\rho, e_1 \eval v_1
	\\
	\rho, e_2 \eval v_2
}
{
	\rho, \tpair{e_1}{e_2} \eval \tpair{v_1}{v_2}
}
\and
\inferrule*
{
	\rho, e \eval \tpair{v_1}{v_2}
}
{
	\rho, \tfst{e} \eval v_1
}
\and
\inferrule*
{
	\rho, e \eval \tpair{v_1}{v_2}
}
{
	\rho, \tsnd{e} \eval v_2
}
\end{ruleblock}
\\[6mm]
\begin{ruleblock}{T::\rho, \mu, M \eval \mu', R}
\inferrule*
{
  	\rho, e \eval v
}
{
	\derivesShade{\tret{e}}{\rho, \mu, \tret{e} \eval \mu, \rret{v}}
}
\and
\inferrule*
{
	\rho, e_1 \eval v_1
	\\
	v_1 = \tclosure{\rho'}{f}{x}{M}
	\\
	\rho, e_2 \eval v_2
	\\
	\derivesShade{T}{\extend{\extend{\rho'}{f}{v_1}}{x}{v_2}, \mu, M \eval \mu', R}
}
{
	\derivesShade{\tappBody{e_1}{e_2}{f}{x}{T}}{\rho, \mu, \tapp{e_1}{e_2} \eval \mu', R}
}
\and
\inferrule*
{
        \rho, e \eval v
}
{
  	\derivesShade{\traise{e}}{\rho,\mu, \traise{e} \eval \mu, \rraise{v}}
}
\and
\inferrule*[right={$\ell \notin \dom(\mu)$}]
{
  	\rho, e \eval v
}
{
	\derivesShade
		{\trefl{\ell}{e}}
		{\rho,\mu,\tref{e} \eval \update{\mu}{\ell}{v}, \rret{\ell}}
}
\and
\inferrule*
{
	\rho,e_1 \eval \ell
	\\
        \rho,e_2 \eval v
}
{
  	\derivesShade
  		{\tassignl{e_1}{\ell}{e_2}}
  		{\rho,\mu, \tassign{e_1}{e_2} \eval \update{\mu}{\ell}{v}, \rret{\tunit}}
}
\and
\inferrule*[right={$\ell \in \dom(\mu)$}]
{
	\rho,e \eval \ell
}
{
	\derivesShade
		{\tderefl{\ell}{e}}
		{\rho,\mu, \tderef{e} \eval \mu, \rret{\mu(\ell)}}
}
\and
\inferrule*
{
  	\derivesShade{T_1}{\rho,\mu, M_1 \eval \mu', \rret{v}}
	\\
	\derivesShade{T_2}{\extend{\rho}{x}{v},\mu', M_2 \eval \mu^\twoPrime, R}
}
{
	\derivesShade
		{\tletSucc{x}{T_1}{T_2}}
		{\rho,\mu, \tlet{x}{M_1}{M_2} \eval \mu^\twoPrime, R}
}
\and
\inferrule*
{
	\derivesShade{T}{\rho,\mu, M_1 \eval \mu', \rraise{v}}
}
{
	\derivesShade{\tletFail{T}}{\rho,\mu, \tlet{x}{M_1}{M_2} \eval \mu', \rraise{v}}
}
\and
\inferrule*
{
	\derivesShade{T_1}{\rho,\mu, M_1 \eval \mu', \rraise{v}}
	\\
	\derivesShade{T_2}{\extend{\rho}{x}{v},\mu', M_2 \eval \mu^\twoPrime, R}
}
{
	\derivesShade{\ttryFail{T_1}{x}{T_2}}{\rho,\mu, \ttry{M_1}{x}{M_2} \eval \mu^\twoPrime, R}
}
\and
\inferrule*
{
	\derivesShade{T_1}{\rho,\mu, M_1 \eval \mu', \rret{v}}
}
{
  	\derivesShade{\ttrySucc{T_1}}{\rho,\mu, \ttry{M_1}{x}{M_2} \eval \mu', \rret{v}}
}
\and
\inferrule*
{
	\rho, e \eval \tinl{v}
	\\
	\derivesShade{T}{\extend{\rho}{x}{v}, \mu, M_1 \eval \mu', R}
}
{
	\derivesShade
		{\tcaseInl{e}{x}{T}{y}}
		{\rho, \mu, \tcase{e}{\tinlC{x}{M_1}}{\tinrC{y}{M_2}} \eval \mu', R}
}
\and
\inferrule*
{
	\rho, e \eval \tinr{v}
	\\
	\derivesShade{T}{\extend{\rho}{x}{v}, \mu, M_2 \eval \mu', R}
}
{
	\derivesShade
		{\tcaseInr{e}{x}{y}{T}}
		{\rho, \mu, \tcase{e}{\tinrC{x}{M_1}}{\tinrC{y}{M_2}} \eval \mu', R}
}
\end{ruleblock}
\caption{Big-step semantics}
\label{fig:semantics}
\end{figure}

For present purposes, we are only concerned with slicing a computation after
it has terminated, so a big-step style of operational semantics seems
appropriate. The evaluation rules are given in Figure~\ref{fig:semantics}.
(The Galois slicing approach was investigated in a small-step style for
$\pi$-calculus by \citet{perera16concur}.) Expression evaluation $\rho, e
\eval v$ says that the expression $e$ evaluates in environment $\rho$ to the
value $v$. Computation evaluation $\rho, \mu, M \eval
\mu', R$, which makes use of expression evaluation, says that computation
$M$ evaluates in environment $\rho$ and store $\mu$ to updated store $\mu'$
and result $R$.  In the latter judgement, we choose to make explicit the
derivation tree $T$ that witnesses the evaluation of $M$ (similarly to how, in
type theory, typed lambda terms are essentially typing derivations); we call
such a proof a
\emph{trace}. To obtain more familiar evaluation rules, it is sufficient to
remove the traces from the judgments; thus we can define
\[
\rho,\mu,M \eval \mu',R \iff \exists T.~T::\rho,\mu,M \eval \mu',R
\]

The trivial computations $\tret{e}$ and $\traise{e}$ evaluate the respective
expressions and return them as a normal or exceptional result respectively.
The evaluation of $\tlet{x}{M}{N}$ corresponds to sequencing: the
subcomputation $M$ is evaluated first and, if it terminates successfully with
$\rret~{v}$, then $N$ is evaluated next (with $v$ substituted for $x$); if $M$
terminates with a failure $\rraise~v$, the whole $\mathtt{let}$ computation
fails with $\rraise~{v}$. The trace forms $\mathtt{let_S}$ and
$\mathtt{let_F}$ correspond to these two possible evaluation outcomes.

The computation $\tref{e}$ chooses a fresh location $\ell$
non-deterministically, extends the store $\mu$ with a new cell at location
$\ell$ containing the result of evaluating $e$, and then returns $\ell$. The
assignment $\tassign{e_1}{e_2}$ evaluates $e_1$ to a location $\ell$ and $e_2$
to a value $v$, updates the cell at $\ell$ with $v$, and returns the unit
value. To evaluate a dereference $\tderef{e}$, we evaluate $e$ to a location
$\ell$, and then return the cell's contents $\mu(\ell)$. For convenience later,
the trace forms for these three computation rules are annotated with the
respective $\ell$ involved in the evaluation.

Function application $e_1~e_2$ combines two pure expressions into an
effectful computation as follows: first $e_1$ is evaluated to a
closure $v_1 = \tclosure{\rho}{f}{x}{M}$, where $M$ is a computation;
then $e_2$ is evaluated to $v_2$. Finally we perform the effectful
evaluation of $M$, where recursive calls $f$ have been replaced by the
closure, and the formal argument $x$ by the actual value $v_2$.

The evaluation of the exception handling $\ttry{M_1}{x}{M_2}$ depends
on the result of the valuation of the subcomputation $M_1$: if it
succeeds with $\rret~{v}$, we simply return this result; if it fails
with $\rraise~{v}$, we proceed to evaluate $M_2$ where $v$ has been
substituted for $x$. The traces $\mathtt{try_S}$ and $\mathtt{try_F}$
correspond to the first and second case respectively.

Finally, case analysis works as usual, taking into account that its
branches are computations, whose effects are triggered after
the substitution of the respective bound variables, similarly to the
function application case.

\begin{theorem}
  If $\rho,e \eval v_1$ and $\rho,e \eval v_2$, then $v_1 = v_2$. 
  \\
  If $T::\rho,\mu,M \eval \mu_1,
  R_1$ and $T::\rho,\mu,M \eval \mu_2, R_2$, then $\mu_1 = \mu_2$ and $R_1
  = R_2$.
\end{theorem}

\subsection{Partial expressions and partial computations}
The language iTML is immediately extended by adding holes to expressions,
computations, and values. This in turn induces the $\sqleq$ relation,
expressing the fact that two terms of the language (expressions, values, and
computations) are structurally equal, save for the fact that some subterms of
the right-hand side term may be matched by holes in the left-hand side term.
The definition of this relation is straightforward, but verbose: we set
$\Hole$ to be the least element and add a congruence rule per constructor.
Figure~\ref{fig:value-leq} illustrates the cases for values, results,
environments, and stores; the cases for expressions and computations are
presented in \appref{prefix}.

Building on partial values, we can view environments and stores as total
functions by defining $\rho(x) = \Hole$ and $\mu(\ell) = \Hole$ whenever $x$
and $\ell$ are not in the domain of $\rho$ and $\mu$. We can then lift
$\sqleq$ pointwise from values to environments and stores. The least
environment, mapping all variables to $\Hole$, is also denoted by $\Hole$; a
similar convention applies to stores.

Meets exist for all pairs of expressions, values, computations, environments
and stores. Furthermore, as we explained in Section~\ref{sec:problem}, we can
define the sets of prefixes of a given language term:
\[
\Prefix{t} = \{ t' : t' \sqleq t \} \qquad t = e, M, v, \rho, \mu 
\]
We can show that in a given prefix set, every pair of terms has a meet
and a join, that is, the sets of partial terms are partonomies for
the corresponding sets of ordinary terms.

\begin{lemma}\label{lem:sqleq-lattice}
For all expressions $e$, computations $M$, values $v$, environments $\rho$,
and stores $\mu$, the sets $\Prefix{e}$, $\Prefix{M}$, $\Prefix{v}$,
$\Prefix{\rho}$ and $\Prefix{\mu}$ form complete lattices with the relation
$\sqleq$.
\end{lemma}

\begin{figure}
  \begin{ruleblock}{v \sqleq v', R \sqleq R', \rho \sqleq \rho', \mu \sqleq \mu'}
\inferrule*
{\strut}
{\Hole \sqleq v}
\and
\inferrule*
{\strut}
{\tunit \sqleq \tunit}
\and
\inferrule*
{\strut}
{\ell \sqleq \ell}
\and
\inferrule*
{v_1 \sqleq v_1' \\
  v_2 \sqleq v_2'}
{\tpair{v_1}{v_2} \sqleq \tpair{v_1'}{v_2'}}
\and
\inferrule*
{v \sqleq v'}
{\tinl{v} \sqleq \tinl{v'}}
\and
\inferrule*
{v \sqleq v'}
{\tinr{v} \sqleq \tinr{v'}}
\and
\inferrule*
{\rho \sqleq \rho' \\
M \sqleq M'}
{\tclosure{\rho}{f}{x}{M} \sqleq \tclosure{\rho'}{f}{x}{M'} }
\and
\inferrule*
{v \sqleq v'}
{\rret~v \sqleq \rret~v'}
\and
\inferrule*
{v \sqleq v'}
{\rraise~v \sqleq \rraise~v'}
\\
\rho \sqleq \rho' \iff \dom(\rho) = \dom(\rho') \land \forall x \in \dom(\rho).~ \rho(x) \sqleq
\rho'(x)
\and
\mu \sqleq \mu' \iff \dom(\mu) = \dom(\mu') \land \forall \ell \in \dom(\mu).~ \mu(x) \sqleq
\mu'(x)
\end{ruleblock}
\caption{Partial value, result, environment, and store prefix relations}
\label{fig:value-leq}
\end{figure}

\subsection{Forward and backward slicing for expressions}

Figure~\ref{fig:slicing-expression} defines the \emph{forward slicing} and
\emph{backward slicing} relations for expressions, which are deterministic and
free of side-effects and recursion. In this situation forward-slicing
degenerates to a form of evaluation extended with a hole-propagation rule. The
judgement $\rho, e \fwd v$ says that partial expression $e$ in partial
environment $\rho$ forward-slices to partial value $v$.

Backward-slicing for expressions is with respect to the original
expression. Suppose $\rho, e \eval v$.  Then for any
$v' \sqsubseteq v$, the judgement $v', e \bwd \rho', e'$ says that
partial value $v'$ backward-slices along expression $e$ to partial
environment $\rho'$ and partial expression $e'$ with
$(\rho', e') \sqsubseteq (\rho, e)$. This must be taken into account
when reading the rules: for example, when we backward-slice $v'$ with
respect to an original expression $x$, the environment
$\extend{\Hole}{x}{v'}$, which maps $x$ to $v'$ and every other variable
in the domain to $\Hole$, is a slice of the original $\rho$.

This consideration proves crucial in the backward slicing rules for pairs. To
slice $\tpair{v_1}{v_2}$ with respect to the original expression
$\tpair{e_1}{e_2}$, we first slice the two component values, obtaining
$\rho_1,e'_1$ and $\rho_2,e'_2$, which are then recombined as $\rho_1 \sqcup
\rho_2,\tpair{e'_1}{e'_2}$. That is, $\rho_1,e_1$ tells us what part of the
environment is needed to force $e_1$ to evaluate to $v_1$ and likewise for
$\rho_1,e_2$, and we combine what we learn about $\rho$ using $\sqcup$.  As
$\rho_1$ and $\rho_2$ are slices of the same original environment, the join
$\rho_1 \sqcup \rho_2$ is guaranteed to exist (Lemma~\ref{lem:sqleq-lattice}).
The rules for slicing functions/closures are the same as given by
\citet{perera12icfp}, and should be considered together with the rule for
slicing function applications in the next section.  We omit rules for
primitive operations, which are handled just as in prior
work~\citep{perera12icfp,acar13jcs}.

\begin{figure}
\begin{ruleblock}{\rho, e \fwd v}
\inferrule*[
]
{
	\strut		
}
{
	\rho, \Hole \fwd \Hole
}
\and
\inferrule*[
]
{
	x \in \dom(\rho)
}
{
	\rho, x\fwd \rho(x)
}
\and
\inferrule*[
]
{
	\strut		
}
{
	\rho, \tunit \fwd \tunit
}
\and
\inferrule*[
]
{
	\strut
}
{
	\rho, \tfun{f}{x}{M} \fwd \tclosure{\rho}{f}{x}{M}
}
\and
\inferrule*
{
	\rho, e \fwd v
}
{
	\rho, \tinl{e} \fwd \tinl{v}	
}
\and
\inferrule*
{
	\rho, e_1 \fwd v_1
	\\
	\rho, e_2 \fwd v_2
}
{
	\rho, \tpair{e_1}{e_2} \fwd \tpair{v_1}{v_2}
}
\and
\inferrule*
{
	\rho, e \fwd \tpair{v_1}{v_2}
}
{
	\rho, \tfst{e} \fwd v_1
}
\and
\inferrule*
{
	\rho, e \fwd \Hole
}
{
	\rho, \tfst{e} \fwd \Hole
}
\and
\inferrule*
{
	\rho, e \fwd \tpair{v_1}{v_2}
}
{
	\rho, \tsnd{e} \fwd v_2
}
\and
\inferrule*
{
	\rho, e \fwd \Hole
}
{
	\rho, \tsnd{e} \fwd \Hole
}
\end{ruleblock}
\\[2mm]
\begin{ruleblock}{v, e \bwd \rho, e'}
\inferrule*[
]
{
	\strut
}
{
	\Hole, e \bwd \Hole, \Hole
}
\and
\inferrule*[
]
{
	v \neq \Hole
}
{
	v, x \bwd \extend{\Hole}{x}{v},  x
}
\and
\inferrule*[
]
{
	\strut
}
{
	\tclosure{\rho}{f}{x}{M}, \tfun{f}{x}{M'} 
	\bwd
	\rho,
	\tfun{f}{x}{M}
}
\and
\inferrule*[
]
{
	\strut
}
{
	\tunit, \tunit \bwd \Hole,  \tunit
}
\and
\inferrule*
{
	v, e \bwd \rho, e'
}
{
	\tinl{v}, \tinl{e} \bwd \rho, \tinl{e'}	
}
\and
\inferrule*
{
	v, e \bwd \rho, e'
}
{
	\tinr{v}, \tinr{e} \bwd \rho, \tinr{e'}	
}
\and
\inferrule*
{
	v_1, e_1 \bwd \rho_1, e_1'
	\\
	v_2, e_2 \bwd \rho_2, e_2'
}
{
	\tpair{v_1}{v_2}, \tpair{e_1}{e_2} \bwd \rho_1 \join \rho_2, \tpair{e_1'}{e_2'}	
}
\and
\inferrule*
{
	\tpair{v}{\Hole}, e \bwd \rho, e'
}
{
	v, \tfst{e} \bwd \rho, \tfst{e'}	
}
\and
\inferrule*
{
	\tpair{\Hole}{v}, e \bwd \rho, e'
}
{
	v, \tsnd{e} \bwd \rho, \tsnd{e'}	
}
\end{ruleblock}
\caption{Forward and backward slicing for expressions}
\label{fig:slicing-expression}
\end{figure}

\begin{lemma}[Forward expression-slicing function]
\label{lem:fwd:functionality:expression}
\item
\begin{enumerate}
\item If $\rho, e \fwd v$ and $\rho, e \fwd v'$ then $v = v'$.
\item Suppose $\rho', e' \eval v'$. If $(\rho, e) \sqleq (\rho',e')$
	there exists $v \sqleq v'$ with $\rho, e \fwd v$.
\end{enumerate}
\end{lemma}

\noindent Given \lemref{fwd:functionality:expression}, we write
$\fwdFs{\rho,e}$ for the function which take any element of $\Prefix{\rho,e}$
to its $\fwd$-image in $\Prefix{v}$.

\begin{lemma}[Meet-preservation]
\label{lem:fwd:meet-preservation:expression}
Suppose $\sigma, e \eval v$.  Then $\fwdFs{\sigma,e}$ preserves $\meet$.
\end{lemma}

Likewise, if we prioritise the use of the first rule (where
$v = \Hole$) over others, backward-slicing determines a deterministic
function, which is total if restricted to any downward-closed subset
of its domain.

\begin{lemma}[Backward expression-slicing function]
\label{lem:bwd:functionality:expression}
\item
\begin{enumerate}
\item If $v, e \bwd \rho, e'$ and $v, e \bwd \rho', e^\twoPrime$ then $(\rho,
e') = (\rho', e^\twoPrime)$.
\item Suppose $\rho, e \eval v$. If $u \sqleq v$ there exists $(\rho', e')
\sqleq (\rho, e)$ such that $u, e \bwd \rho', e'$.
\end{enumerate}
\end{lemma}

\noindent Given \lemref{bwd:functionality:expression}, we write
$\bwdFs{\rho,e}$ for the function which takes any element of $\Prefix{v}$ to
its $\bwd$-image in $\Prefix{\rho,e}$. The functions $\fwdFs{\rho,e}$ and
$\bwdFs{\rho,e}$ form a Galois connection. This is essentially a special case
of the Galois connection defined by \citet{perera12icfp}, where the traces
associated with the expression forms are simply the expressions themselves.
 
\begin{theorem}[Galois connection for expression slicing]
\label{thm:galois-connection:expression}
Suppose $\rho, e \eval v$.  Then $ \bwdFs{\rho,e}\dashv\fwdFs{\rho,e} $.
\end{theorem}

\section{Slicing for references and exceptions}
\label{sec:coarse-slicing}

In previous work on slicing for (pure) TML, slicing was initially focused on
programs only, with forward slicing taking a program as its input, and
backward slicing producing a slice of the original program. As program slices
proved inadequate to explaining the behaviour of more complex code, a separate
trace slicing procedure was also defined that produced a slice of the
execution trace. Trace slicing has also proved useful in subsequent work on the
$\pi$-calculus, where it is needed to determine a specific concurrent
behaviour in an inherently nondeterministic semantics \citep{perera16concur}.

Evaluation for iTML is also non-deterministic in that it models the allocation
of a new reference by chosing any unused store location. This nondeterminism
is weak, in the sense that any two executions from the same initial state
yield isomorphic results (up to the permutation of newly allocated locations).
It is technically possible to determinise allocation and define forward
slicing as a total function in the presence of references. However, if forward
slicing only has access to the expression and input, it must be extremely
conservative when forward slicing a hole: since we do not know the locations
which were written at run-time by the missing expression, we must
conservatively assume that any location may have changed, erasing the whole
store. This, in turn, forces backward slicing to retain all of the write
operations in the program, even those that seem to have nothing to do with the
slicing criterion.

While adding exceptions to a pure language does not necessarily introduce
nondeterminism, exceptions also complicate the picture for slicing
considerably: if we replace a subexpression with a hole, then (in the absence
of information about what happened at runtime) it is impossible to know
whether that expression terminated normally or raised an exception.  This
means that we may be forced to retain many parts of the program solely to
ensure that we can always be certain whether or not an exception was raised.

Since trace information has proven useful for implementing backward
slicing even for pure programs, seems well-motivated for dealing with
exceptions and references, and is in any case necessary for other
features such as concurrency or true nondeterminism, we accordingly
propose the following generalisation of Galois slicing, which takes
explicit account of traces.  Specifically, we consider \emph{tracing}
computations $C \subseteq X \times T \times Y$, where $T$ is some set
of traces that describe what happened in a given run of $C$, such that
for any $(x,t)$ there is a unique $y$ such that $(x,t,y) \in C$.  We
assume $X,Y$ and $T$ are equipped with partonomies so that
$\Prefix{x}, \Prefix{y}$ and $\Prefix{t}$ are complete lattices for
any $x\in X, y \in Y$ and $t \in T$.  Given $(x,t,y) \in C$, we define
a meet-preserving function
$\fwdF : \Prefix{x} \times \Prefix{t} \to \Prefix{y}$ that computes as
much as possible of $y$ given the partial information about the input
in $x$ and about the trace in $t$.  Then a lower adjoint
$\bwdF : \Prefix{y} \to \Prefix{x}\times \Prefix{t}$ is uniquely
determined, and produces the least partial input and partial trace
that suffices to recompute a given partial input.

Traces, like iTML computations, are made partial by adding holes. However,
rather than using a single, fully undefined trace $\Box$ that could stand for
an entirely arbitrary evaluation, providing no information about its result,
the parts of store that have been written, or even whether the computation
succeeded or raised an exception, we provide annotated trace holes allowing
for a less draconian slicing.

An annotated trace hole will be written $\thole{\Ell}{\kay}$, where $\Ell$ is
the set of store locations written by the otherwise unknown trace and $\kay$
is the outcome. Unlike the unannotated hole $\Hole$ used in the pure setting,
annotated holes retain information about the effects and outcome of the
computation. Thus, even though annotated holes do not say exactly how a
computation evaluated and what its result (or exception) value was, they still
disclose information about its side effects.

The $\sqleq$ relation for traces is defined analogously to $\sqleq$ for
values, except that there is no universal least trace $\Hole$. Rather, for any
trace $T$, the hole $\thole{\Ell}{\kay}$ is the least element of $\Prefix{T}$
where $\Ell =
\writes(T)$ and $\kay = \res(T)$:
\begin{smathpar}
\inferrule*
{
	\writes(T) = \Ell
	\\
	\res(T) = \kay
}
{
	\thole{\Ell}{\kay} \sqleq T
}
\end{smathpar}

\vspace{-10pt}
\noindent The auxiliary operations $\writes(T)$ and $\res(T)$ are defined in
Figure~\ref{fig:writes}. The former computes the set of locations allocated or
updated by $T$; the latter indicates whether $T$ returned normally or raised
an exception. Thus $\thole{\Ell}{\kay}$ represents the full erasure of a
computation that writes to locations in $\Ell$ and returns as described by
$\kay$. The $\sqleq$ relation is simply the compatible closure of the above
rule; the full definition is given in
\appref{prefix}.

\begin{lemma}
\label{lem:sqleq-lattice-trace}
For all traces $T$, the set $\Prefix{T}$ forms a complete lattice with the relation $\sqleq$.
\end{lemma}

\begin{figure}
\small
\begin{minipage}{7cm}\begin{align*}
\writes(\thole{\Ell}{\kay})
&=
\Ell
\\
\writes(\tret{e})
&=
\varnothing
\\
\writes(\tletFail{T_1})
&=
\writes(T_1)
\\
\writes(\tletSucc{x}{T_1}{T_2})
&=
\writes(T_1) \cup \writes(T_2)
\\
\writes(\tappBody{e_1}{e_2}{f}{x}{T})
&=
\writes(T)
\\
\writes(\tcaseInl{e}{x}{T}{y})
&=
\writes(T)	
\\
\writes(\tcaseInr{e}{x}{y}{T})
&=
\writes(T) 
\\
\writes(\traise{e})
&=
\varnothing
\\
\writes(\ttrySucc{T_1})
&=
\writes(T_1)
\\
\writes(\ttryFail{T_1}{x}{T_2})
&=
\writes(T_1) \cup \writes(T_2)
\\
\writes(\trefl{\ell}{e})
&=
\set{\ell}
\\
\writes(\tassignl{e_1}{\ell}{e_2})
&=
\set{\ell}
\\
\writes(\tderefl{\ell}{e})
&=
\emptyset
\end{align*}
\end{minipage}
~
\begin{minipage}{7cm}
\begin{align*}
\res(\thole{\Ell}{\kay}) 
&= \kay
\\
\res(\tret{e})
&=
\rret
\\
\res(\tletFail{T_1})
&=
\rraise
\\
\res(\tletSucc{x}{T_1}{T_2})
&=
\res(T_2)
\\
\res(\tappBody{e_1}{e_2}{f}{x}{T})
&=
\res(T)
\\
\res(\tcaseInl{e}{x}{T}{y})
&=
\res(T)
\\
\res(\tcaseInr{e}{x}{y}{T})
&=
\res(T)
\\
\res(\traise{e})
&=
\rraise
\\
\res(\ttrySucc{T_1})
&=
\rret
\\
\res(\ttryFail{T_1}{x}{T_2})
&=
\res(T_2)
\\
\res(\trefl{\ell}{e})
&=
\rret
\\
\res(\tassignl{e_1}{\ell}{e_2})
&=
\rret
\\
\res(\tderefl{\ell}{e})
&=
\rret
\end{align*}
\end{minipage}
\caption{Set of store locations $\writes(T)$ written to by $T$ and
  outcome $\res(T)$ of $T$.}
\label{fig:writes}
\end{figure}

We will define both forward and backward slicing as judgments, and show that
when $T:: \rho,\mu_1,e \eval \mu_2,R$ we can define a Galois connection
$\bwdF\dashv\fwdF$ between $\Prefix{\rho,\mu_1,e,T}$ and $\Prefix{\mu_2,R}$.
We first motivate our definition of forward slicing, then outline its
properties, particularly meet-preservation. We then present the rules for
backward slicing.  Although backward slicing is uniquely determined by forward
slicing, we give rules that show how to compute backward slicing more
efficiently than the naive approach.  The main idea, as in the pure case, is
to use the trace structure to guide backward slicing.  Nevertheless, due to
the presence of side-effects and exceptions, there are a number of subtleties
that do not arise in the pure case. The rules we give will make use of an
operation for partial store erasure, which takes a store and a set of
locations $\Ell$, and returns a copy of that store with all locations in
$\Ell$ replaced by hole.

\begin{definition}
\label{def:erasure}
For partial store $\mu$ and set of locations $\Ell$, the \emph{store erasure}
operation $\mu \antirestrict \Ell$ is defined as follows:
\begin{align*}
\mu \antirestrict \Ell
&=
\update{\mu}{\ell}{\Hole \mid \ell \in \Ell}
\end{align*} 
\end{definition}

\subsection{Forward slicing}

In a pure language, a subexpression whose value is not needed by the rest of
the computation can be sliced away safely, because we know that any other
expression evaluated in its place will not raise an exception or have
side-effects.  However, when side-effects or exceptions are added to the
picture, we need to be more careful when slicing subexpressions whose values
were not needed, because the expression may have had side-effects on store
locations, or the fact that the expression raised an exception may have been
important to the control flow of the program.  For this reason, we allow
forward slicing to consult the trace, so that when information about reference
side-effects or control flow is not present in the partial program, we can
recover it from the trace.  Thus, forward slicing for computations is with
respect to a partial trace $T$ which determinises allocations and enables
forward-slicing of store effects.

Suppose that $\derives{T}{\rho, \mu_1, M \eval \mu_2, R}$.  The
forward slicing judgement
$\rho', \mu_1', M', T' \fwd \mu_2', R'$ takes a partial expression
$e' \sqleq e$, partial environment $\rho'\sqleq \rho$, partial store
$\mu_1'\sqleq \mu_1$, and partial trace $T' \sqleq T$, and should
produce an updated store $\mu_2'\sqleq \mu_2$ and result
$R' \sqleq R$. 

Figures~\ref{fig:slicing-fwd-coarse1} and~\ref{fig:slicing-fwd-coarse2} define the forward slicing rules
for computations, which are named for convenience.  The significance
of $T'$ being partial is that forward slicing may be performed with a
computation that has already been sliced, so that the forward slicing
rules induce a total function from partial inputs
$\Prefix{\rho,\mu_1,M,T}$ to partial outputs $\Prefix{\mu_2,R}$. The
rules for forward slicing deserve explanation, since they embed
important design decisions regarding how partial inputs can be used to
compute partial outputs, which in turn affect the definition of
backward slicing.

\begin{figure}
\begin{ruleblock}{\rho, \mu, M, T \fwd \mu', R}
\inferrule[F-Trace$\Hole$]
{
	\strut
}
{
   \rho, \mu, M, \thole{\Ell}{\kay}
   \fwd
   \mu \antirestrict \Ell, \kay~\Hole
}
\and
\inferrule[F-Comp$\Hole$]
{
  \Ell = \writes(T)\\
\kay = \res(T)
}
{
	\rho, \mu, \Hole, T 
	\fwd 
	\mu\antirestrict \Ell, \kay~\Hole
}
\and
\inferrule[{F-Ret}
]
{
	\rho, e \fwd v
}
{
	\rho, \mu, \tret{e}, \tret{e'} 
	\fwd 
	\mu, \rret{v}
}
\and
\inferrule[{F-Let}
]
{
	\rho, \mu, M_1, T_1 \fwd \mu', \rret{v}
	\\
	\extend{\rho}{x}{v}, \mu', M_2, T_2 \fwd \mu^\twoPrime, R
}
{
	\rho, \mu, \tlet{x}{M_1}{M_2}, \tletSucc{x}{T_1}{T_2}
	\fwd
	\mu^\twoPrime, R
}
\and
\inferrule[{F-LetFail}
]
{
	\rho, \mu, M_1, T_1 \fwd \mu', \rraise{v}
}
{
	\rho, \mu, \tlet{x}{M_1}{M_2}, \tletFail{T_1}
	\fwd
	\mu', \rraise{v}
}
\and
\inferrule[F-CaseL]
{
	\rho, e \fwd \tinl{v}
	\\
	\extend{\rho}{x}{v}, \mu, M_1, T \fwd \mu', R
}
{
	\rho, \mu, \tcase{e}{\tinlC{x}{M_1}}{\tinrC{y}{M_2}}, \tcaseInl{e'}{x}{T}{y} 
	\fwd 
	\mu', R
}
\and
\inferrule[F-CaseR]
{
	\rho, e \fwd \tinr{v}
	\\
	\extend{\rho}{y}{v}, \mu, M_2, T \fwd \mu', R
}
{
	\rho, \mu, \tcase{e}{\tinlC{x}{M_1}}{\tinrC{y}{M_2}}, \tcaseInr{e'}{x}{y}{T} 
	\fwd 
	\mu', R
}
\and
\inferrule[F-CaseL$\Hole$]
{
	\rho, e \fwd \Hole
	\\
	\Ell = \writes(T) \\
        \kay = \res(T)
}
{
  	\rho, \mu, \tcase{e}{\tinlC{x}{M_1}}{\tinrC{y}{M_2}}, \tcaseInl{e'}{x}{T}{y} 
	\fwd 
	\mu\antirestrict \Ell, \kay~\Hole
}
\and
\inferrule[F-CaseR$\Hole$]
{
	\rho, e \fwd \Hole
	\\
	\Ell = \writes(T) \\
        \kay = \res(T)
}
{
  	\rho, \mu, \tcase{e}{\tinrC{x}{M_1}}{\tinrC{y}{M_2}}, \tcaseInr{e'}{x}{y}{T} 
	\fwd 
	\mu\antirestrict \Ell, \kay~\Hole
}
\and
\inferrule[F-App
]
{
	\rho, e_1 \fwd v_1
	\\
    v_1 = \tclosure{\rho'}{f}{x}{M}
    \\
	\rho, e_2 \fwd v_2
	\\
	\extend{\extend{\rho'}{f}{v_1}}{x}{v_2}, \mu, M, T \fwd \mu', R
}
{
	\rho, \mu, \tapp{e_1}{e_2}, \tappBody{e_1'}{e_2'}{f}{x}{T} 
	\fwd 
	\mu', R
}
\and
\inferrule[F-App$\Hole$
]
{
   \rho, e_1 \fwd \Hole
   \\
   \Ell = \writes(T) \\
        \kay = \res(T)
}
{
   \rho, \mu, \tapp{e_1}{e_2}, \tappBody{e_1'}{e_2'}{f}{x}{T} 
   \fwd 
   \mu \antirestrict \Ell, \kay~\Hole
}
\end{ruleblock}
\caption{Forward slicing for computations: holes, let-bindings, cases and
  function applications}
\label{fig:slicing-fwd-coarse1}
\end{figure}
\begin{figure}
\begin{ruleblock}{\rho, \mu, M, T \fwd \mu', R}
\inferrule[F-Raise
]
{
	\rho, e\fwd v
}
{
	\rho, \mu, \traise{e}, \traise{e'} 
	\fwd 
	\mu, \rraise{v}
}
\and
\inferrule[F-Try
]
{
	\rho,\mu, M_1, T_1 \fwd \mu', \rret{v}
}
{
	\rho, \mu', \ttry{M_1}{x}{M_2}, \ttrySucc{T_1}
	\fwd
	\mu', \rret{v}
}
\and
\inferrule[F-TryFail
]
{
	\rho, \mu, M_1, T_1 \fwd \mu', \rraise{v}
	\\
	\extend{\rho}{x}{v}, \mu', M_2, T_2 \fwd \mu^\twoPrime, R
}
{
	\rho, \mu, \ttry{M_1}{x}{M_2}, \ttryFail{T_1}{x}{T_2}
	\fwd
	\mu^\twoPrime, R
}
\and
\inferrule[F-Ref
]
{
	\rho, e \fwd v	
}
{
	\rho, \mu, \tref{e}, \trefl{\ell}{e'}
	\fwd
	\update{\mu}{\ell}{v}, \rret{\ell}
}
\and
\inferrule[F-Assign
]
{
	\rho, e_1 \fwd \ell
	\\
	\rho, e_2 \fwd v
}
{
	\rho, \mu, \tassign{e_1}{e_2}, \tassignl{e_1'}{\ell}{e_2'} 
	\fwd
	\update{\mu}{\ell}{v}, \rret{\tunit}
}
\and
\inferrule[F-Assign$\Hole$
]
{
	\rho, e_1 \fwd \Hole
}
{
	\rho, \mu, \tassign{e_1}{e_2}, \tassignl{e_1'}{\ell}{e_2'} 
	\fwd
	\update{\mu}{\ell}{\Hole}, \rret{\Hole}
}
\and
\inferrule[F-Deref
]
{
	\rho, e \fwd \ell
}
{
	\rho, \mu, \tderef{e}, \tderefl{\ell}{e'} 
	\fwd 
	\mu, \rret{\mu(\ell)}
}
\and
\inferrule[F-Deref$\Hole$
]
{
	\rho, e \fwd \Hole
}
{
	\rho, \mu, \tderef{e}, \tderefl{\ell}{e'} 
	\fwd 
	\mu, \rret{\Hole}
}
\end{ruleblock}
\caption{Forward slicing for computations: exceptions and references}
\label{fig:slicing-fwd-coarse2}
\end{figure}

The first two rules \rulename{F-Trace$\Hole$} and \rulename{F-Comp$\Hole$} are
the most important.  Recall that a trace hole $\thole{\Ell}{\kay}$ is
annotated with the set of locations $\Ell$ written to by the trace.  The
\rulename{F-Trace$\Hole$} rule covers the case where
$T = \thole{\Ell}{\kay}$.  This means that we know only that while executing
$M$, the original computation wrote to locations in $\Ell$ and eventually had
outcome $\kay$, but we have no other information about what values were
written to the locations in $\Ell$ or what value was returned.  Thus, we have
little choice but to erase the locations in $\Ell$ in the store and yield
result $\kay~\Hole$, that is, we know that evaluation returned with outcome
$\kay$, but nothing about the value returned.

The \rulename{F-Comp$\Hole$} rule covers the case where $M = \Hole$.  In this
case, we rely on information in the trace to approximate the downstream
effects on the store.  We could, in principle, use $T$ to continue
recomputing, but we choose not to, since the goal of forward slicing is to
show how much output can be computed from the program $M$. Instead, we use the
auxiliary function $\writes$ to find the set of locations written to by $T$,
and the erasure operation $\mu \antirestrict \Ell$ defined in
Figure~\ref{fig:writes}, to erase all locations written to by $T$, and return
$\kay~\Hole$ where $\kay = \res(T)$.

Observe that the \rulename{F-Comp$\Hole$} and
\rulename{F-Trace$\Hole$} rules overlap in the case
$M = \Hole, T = \thole{\Ell}{\kay}$, and in this case their behaviour
is identical because $\writes(\thole{\Ell}{\kay}) = \Ell$ and
$\res(\thole{\Ell}{\kay}) = \kay$.

Many of the rules are the same (modulo minor syntactic differences) as
the tracing evaluation rules. The rest of the rules handle situations
where a partial expression or computation forward slices to $\Hole$. We
discuss two of these rules, \rulename{F-Assign$\Hole$} and \rulename{F-Deref$\Hole$}, in detail.
The \rulename{F-Assign$\Hole$}  rule deals with the possibility that while
evaluating $e_1 := e_2$, the first subexpression evaluates to $\Hole$.  In
this case, we return $\Hole$, and update the store so that $\mu(\ell) = \Hole$
as well, where $\ell$ is the updated location recorded in the trace.  This
rule illustrates the benefits of using the trace: without knowing $\ell$, we
would be forced to conservatively set the whole store to $\Hole$, since we
would have no way to be sure which location was updated.  Nevertheless, we do
not continue evaluating using the expressions $e_1',e_2'$ stored in the trace;
they are ignored in forward slicing, but are necessary for backward slicing.

The \rulename{F-Deref$\Hole$} rule deals with the possibility that when evaluating
$\bang e$, the subexpression $e$ evaluates to $\Hole$.  Again, we
cannot be certain what the value of $\mu(\ell)$ is so we simply return
$\Hole$.  Here the subscript $\ell$ is not needed, but again it will be
needed for backward slicing.

The remaining rules \rulename{F-CaseL$\Hole$ }, \rulename{F-CaseR$\Hole$},
and \rulename{F-App$\Hole$} deal with the cases for case expressions or
function applications in which the first argument evaluates to an
unknown value or outcome $\Hole$.  In the case expression and function
application cases, since we cannot proceed with evaluation, we proceed
as in the previous case.  In the other case, such as let-expressions
or try-blocks, note that it is impossible for the outcome of the first
subcomputation to be unknown, since we do not allow unknown outcomes
$\Hole$.

Forward slicing is deterministic, and total when restricted to
downward-closed subsets of its domain. In particular in the
$\tref{\param}$ rules, $\ell$ is fixed by the fact that we can consult
the trace $\trefl{\ell}{e'} $ of
$\rho, \mu,\tref{e} \eval \update{\mu'}{\ell}{v}$ which records the
already-chosen location of $\ell$.  Without the trace argument,
forward slicing would not be deterministic, just as ordinary evaluation
is not.

\begin{lemma}[Forward slicing function]
\label{lem:fwd:functionality:computation}
\item
\begin{enumerate}
\item If $\rho, \mu_1, M, T \fwd \mu_2, R$ and $\rho, \mu_1, M, T \fwd
\mu_2', R'$ then $(\mu_2, R) = (\mu_2',R')$.
\item Suppose $\derives{T}{\rho, \mu_1, M \eval \mu_2, R}$. If $(\rho',
	\mu_1', M', T') \sqleq (\rho, \mu_1, M, T)$ there exists $(\mu_2', R') \sqleq
	(\mu_2, R)$ with $\rho', \mu_1', M', T' \fwd \mu_2', R'$.
\end{enumerate}
\end{lemma}

\noindent Given \lemref{fwd:functionality:computation}, we write
$\fwdFs{T}$ for the function which takes any element of
$\Prefix{\rho', \nu, M', T}$ to its forward image in
$\Prefix{\nu', R'}$.  (We write $\fwdFs{T}$ instead of
$\fwdFs{\rho,\mu,M,T}$ because $T$ provides enough information to
determinise evaluation.)

\begin{lemma}[Meet-preservation for $\fwdFs{T}$]
\label{lem:fwd:meet-preservation:computation}
Suppose $\derives{T}{\rho, \mu_1, M \eval \mu_2, R}$ and $x,x'
\sqleq (\rho, \mu_1, M, T)$. Then $\fwdFs{T}(x\meet x') = \fwdFs{T}(x)
\meet \fwdFs{T}(x')$.
\end{lemma}

\subsection{Backward slicing}

We will define backward slicing inductively using rules for the
judgment $\mu,R,T \bwd \rho,\mu',M,U$, which can be read as ``To
produce partial output store $\mu$, partial result $R$, and partial
trace $T$, the input environment $\rho$, input store $\mu'$, program
$M$ and trace $U$ are required''.  The first two arguments $\mu$ and
$R$ constitute the \emph{slicing criterion}, where $\mu$ allows us to
specify what parts of the output store are of interest, and $T$ is a
trace of the computation (obtained initially from tracing evaluation
of the program.)

\begin{figure}
\begin{ruleblock}{\mu, R, T \bwd \rho, \mu', M, U}
\inferrule[B-Slice$\Hole$
]
{
	\Ell = \writes(T)
	\\
    \mu \antirestrict \Ell = \mu
}
{
	\mu, \kay~\Hole, T
	\bwd 
	\Hole, \mu, \Hole, \thole{\Ell}{\kay}
}
\and
\inferrule[B-Ret
]
{
	v, e \bwd \rho,  e'
}
{
	\mu, \rret{v}, \tret{e} 
	\bwd
	\rho, \mu, \tret{e'}, \tret{e'}
}
\and
\inferrule[B-Let
]
{
	\mu, R, T_2 \bwd \extend{\rho_2}{x}{v}, \mu', M_2, U_2
	\\
	\mu', \rret{v}, T_1 \bwd \rho_1, \mu^\twoPrime, M_1, U_1
}
{
	\mu, R, \tletSucc{x}{T_1}{T_2} 
	\bwd
	\rho_1 \join \rho_2, \mu^\twoPrime, \tlet{x}{M_1}{M_2}, \tletSucc{x}{U_1}{U_2}
}
\and
\inferrule[B-LetFail]
{
	\mu, \rraise{v}, T_1 \bwd \rho, \mu', M_1, U_1
}
{
	\mu, \rraise{v}, \tletFail{T_1}
	\bwd
	\rho, \mu', \tlet{x}{M_1}{\Hole}, \tletFail{U_1}
}
\and
\inferrule[B-CaseL]
{
	\mu, R, T \bwd \extend{\rho}{x}{v}, \mu', M_1, U
	\\
	\tinl{v}, e \bwd \rho', e'
}
{
	\mu, R, \tcaseInl{e}{x}{T}{y} 
	\bwd 
	\rho \join \rho', \mu', \tcase{e'}{\tinlC{x}{M_1}}{\tinrC{y}{\Hole}}, \tcaseInl{e'}{x}{U}{y}
}
\and
\inferrule[B-CaseR]
{
	\mu, R, T \bwd \extend{\rho}{y}{v}, \mu', M_2, U
	\\
	\tinr{v}, e \bwd \rho', e'
}
{
	\mu, R, \tcaseInr{e}{x}{y}{T}
	\bwd 
	\rho \join \rho', \mu', \tcase{e'}{\tinrC{x}{\Hole}}{\tinrC{y}{M_2}}, \tcaseInr{e'}{x}{y}{U}
}
\and
\inferrule[B-App]
{
	\mu, R, T 
	\bwd 
	\extend{\extend{\rho}{f}{v_1}}{x}{v_2}, \mu', M, U
    \\
	v_2, e_2 \bwd \rho_2, e_2'
	\\
	v_1 \join \tclosure{\rho}{f}{x}{M}, e_1 \bwd \rho_1, e_1'
}
{
	\mu, R, \tappBody{e_1}{e_2}{f}{x}{T} 
	\bwd 
	\rho_1 \join \rho_2, \mu', \tapp{e_1'}{e_2'}, \tappBody{e_1'}{e_2'}{f}{x}{U} 
}
\and
\inferrule[B-Raise
]
{
	v, e \bwd \rho, e'
}
{
	\mu, \rraise{v}, \traise{e} \bwd \rho, \mu, \traise{e'}, \traise{e'}
}
\and
\inferrule[B-TryFail
]
{
	\mu, R, T_2 \bwd \extend{\rho_1}{x}{v}, \mu', M_2, U_2
	\\
	\mu', \rraise{v}, T_1 \bwd \rho_2, \mu^\twoPrime, M_1, U_1
}
{
	\mu, R, \ttryFail{T_1}{x}{T_2} 
	\bwd
	\rho _1\join \rho_2, \mu^\twoPrime, \ttry{M_1}{x}{M_2}, \ttryFail{U_1}{x}{U_2}
}
\and
\inferrule[B-Try
]
{
	\mu, \rret{v}, T_1 \bwd \rho, \mu', M_1, U_1
}
{
	\mu, \rret{v}, \ttrySucc{T_1} 
	\bwd
	\rho, \mu', \ttry{M_1}{x}{\Hole}, \ttrySucc{U_1}
}
\and
\inferrule[B-Ref
]
{
	\mu(\ell), e \bwd \rho, e'
}
{
	\mu, \rret{v}, \trefl{\ell}{e} 
	\bwd 
	\rho, \update{\mu}{\ell}{\Hole}, \tref{e'}, \trefl{\ell}{e'}
}
\and
\inferrule[B-Assign
]
{
	\mu(\ell), e_2 \bwd \rho_2, e_2'
	\\
	\ell, e_1 \bwd \rho_1, e_1'
}
{
	\mu, \rret{v}, \tassignl{e_1}{\ell}{e_2} 
	\bwd 
	\rho_1 \join \rho_2, 
	\update{\mu}{\ell}{\Hole}, \tassign{e_1'}{e_2'}, \tassignl{e_1'}{\ell}{e_2'}
}
\and
\inferrule[B-Deref
]
{
	\ell, e \bwd \rho, e'
}
{
	\mu, \rret{v}, \tderefl{\ell}{e} 
	\bwd 
	\rho, \mu \join \update{}{\ell}{v}, \tderef{e'}, \tderefl{\ell}{e'}
}
\end{ruleblock}
\caption{Backward slicing for computations }
\label{fig:slicing-bwd-coarse}
\end{figure}

Figure~\ref{fig:slicing-bwd-coarse} defines backward slicing for
computations.  We explain these rules in greater detail, because
there are a number of subtleties relative to the rules for slicing
pure programs.  

The \rulename{B-Slice$\Hole$} rule is applied preferentially whenever
possible, to avoid a profusion of straightforward but verbose side-conditions.
This rule says that if the return value of the slicing criterion is not needed
\emph{and} none of the locations $\Ell$ written to by $T$ are needed (i.e.
$\mu
\antirestrict \Ell = \mu$), then we return the empty environment $\Hole$,
unchanged store $\mu$, empty program $\Hole$, and hole trace
$\thole{\Ell}{\kay}$ recording the write set and outcome of $T$.  The idea
here is that we are allowed to slice away information that contributed only to
the outcome of a computation that returns normally, as long as the result
value or side effects of the computation are not needed.  Thus, the annotated
trace hole $\thole{\Ell}{\kay}$ records just enough information about $T$ to
allow us to approximate its behaviour during forward slicing.

The rule \rulename{B-Ret} for slicing $T = \tret{e}$ is
straightforward; we use the expression slicing judgment. For
let-binding, there are two rules: \rulename{B-Let} for
$T = \tletSucc{x}{T_1}{T_2}$ when the first subexpression returns, and
\rulename{B-LetFail} for $T = \tletFail{T}$ when the first
subexpression raises an exception.  In the first case, we slice $T_2$
with respect to the result of the computation.  This yields an
environment of the form $\rho[x\mapsto v]$, where $v$ shows what part
of the value of $x$ was required in $T_2$.  We then slice $T_1$ with
respect to $\rret~{v}$.  The partial environments $\rho_1$ and
$\rho_2$ resulting from slicing the subtraces are joined, while the
store $\mu$ is threaded through the slicing judgments for $T_2$ and
$T_1$.  The rule for $T=\tletFail{T'}$ simply slices $T'$ with respect
to the result $R$ (which may be $\Hole$ or $\rraise~v$).

The rules for slicing case expressions \rulename{B-CaseL},
\rulename{B-CaseR} and applications \rulename{B-App} are similar to
those for the corresponding constructs in pure TML; we briefly
summarize them.  When we slice $T = \tcaseInl{e}{x}{T'}{y}$, we slice
$T'$ with respect to the result, and obtain the value $v$ showing what
part of $x$ is needed; we then slice $e$ with respect to $\tinl{v}$.
The rule for $\tcaseInr{e}{x}{y}{T'}$ is symmetric.  Finally, for
application traces $T = \tappBody{e_1}{e_2}{f}{x}{T'}$, we slice $T'$
with respect to the outcome, and obtain from this $v_1$ and $v_2$
which show how much of the function and argument were needed for the
recursive call.  We also obtain $\rho$ which shows what other values
in the closure were needed and $M$ which shows what part of the
function body was needed.  The argument expression $e_2$ is then
sliced with respect to $v_2$ and the function expression $e_1$ is
sliced with respect to $v_1 \sqcup \tclosure{\rho}{f}{x}{M}$.  (The
\rulename{B-App} rule illustrates an additional benefit of combining
program and trace slicing in a single judgment; \citet{perera12icfp} treated
program slicing and trace slicing separately, which made it necessary to
traverse the subtrace $T'$ twice in order to perform trace slicing).

The rules for raising exceptions \rulename{B-Raise}, and for dealing
with $\mathtt{try}$ \rulename{B-Try}, \rulename{B-TryFail}, are
exactly symmetric to the rules for returning normally and for
let-binding.

The rules for references deserve careful examination.  For reference cell
creation, in rule \rulename{B-Ref} we slice the expression $e$ with respect to
the value of $\mu(\ell)$, where $\ell$ is the location recorded in the trace,
and we update the store to map $\ell$ to $\Hole$ since $\ell$ is not allocated
before the reference is created.  The result $R$ is irrelevant in this rule,
since rerunning the reference expression will fully restore the return value
$\ell$.

For assignment, the rule \rulename{B-Assign} 
slices $e_2$ with
respect to $\mu(\ell)$ and we slice $e_1$ with respect to $\ell$
itself.  Finally, we update the store so that $\ell$ is mapped to
$\Hole$; this is necessary because we have no way of knowing the value
of $\ell$ before the assignment, and in any case it should be removed
from the slicing criterion until any earlier reads from $\ell$ are
considered. 
Finally, for dereferencing the rule \rulename{B-Deref} handles the
case where we slice with respect to a known return value $\rret~v$.
In this case, we can assume $v \neq \Hole$, since otherwise an earlier
rule would apply.  Thus we slice $e$ with respect to $\ell$ and we add
$v$ to the slicing criterion $\mu(\ell)$.

Because of the prioritisation of the first rule, backward slicing is
deterministic, and total for downward-closed subsets of its domain. Note that
this preference for the first rule means that the other rules will only be
used when either the value part of the result is not $\Hole$, or there are
locations $\ell \in
\writes(T)$ such that $\mu(\ell) \neq \Hole$.  In particular, rules
\rulename{B-Ref} and
\rulename{B-Assign} will only be used when the either the returned
value ($\ell$ or $\Hole$) or the value of the location $\ell$ that is created
or assigned is part of the slicing criterion, i.e. $\mu(\ell) \neq \Hole$.

\begin{lemma}[Backward slicing function]
\label{lem:bwd:functionality:computation}
\item 
\begin{enumerate}
	\item If $\mu, R, T \bwd \rho, \mu', M, U$ and $\mu, R, T \bwd \rho',
\mu^\twoPrime, M', U'$ then $(\rho, \mu', M, U) = (\rho', \mu^\twoPrime, M',
U')$.
	\item Suppose $\derives{T}{\rho', \nu, M' \eval \nu', R'}$. If $(\mu, R)
	\sqleq (\nu', R')$ there exists $(\rho, \mu', M, U)
	\sqleq (\rho', \nu, M', T)$ such that $\mu, R, T \bwd \rho, \mu', M, U$.
\end{enumerate}
\end{lemma}

\noindent Given \lemref{bwd:functionality:computation}, we write $\bwdFs{T}$
for the function which takes any element of $\Prefix{\nu', R'}$ to its
$\bwd$-image in $\Prefix{\rho', \nu, M', T}$. It computes the lower adjoint of the forward
slicing function for a given computation.

\begin{theorem}[Galois connection for a computation]
\label{thm:galois-connection:computation}
\item Suppose $\derives{T}{\rho', \nu, M' \eval \nu', R'}$.
\begin{enumerate}
	\item If $(\rho, \mu, M, U) \sqleq (\rho', \nu, M', T)$ then
	$\bwdFs{T}(\fwdFs{T}(\rho, \mu, M, U)) \sqleq (\rho, \mu, M, U)$.
	\item If $(\mu, R) \sqleq (\nu', R')$ then $\fwdFs{T}(\bwdFs{T}(\mu, R))
	\sqgeq (\mu, R)$.
\end{enumerate}
\end{theorem}

\noindent Analogously to the Galois connection for an expression,
\thmref{galois-connection:computation} implies that $\bwdFs{T}$ preserves
joins (and is therefore monotonic).

\begin{lemma}[Join-preservation for $\bwdFs{T}$]
\label{lem:bwd:join-preservation:computation}
Suppose $\derives{T}{\sigma, \nu, M \eval \nu', S}$ and $(\mu, R), (\mu',
R') \sqleq (\nu', S)$. Then $\bwdFs{T}(\mu \join \mu', R \join R') =
\bwdFs{T}(\mu, R) \join \bwdFs{T}(\mu', R')$.
\end{lemma}

\section{Arrays, sequential composition, and loops}
\label{sec:extensions}

Any self-respecting imperative language includes mutable arrays,
sequential composition, and loops.  In this section we sketch how they
can be added to our framework.  

We first consider the following extension to the computations and traces to
accommodate arrays:
\begin{eqnarray*}
M& ::=& \cdots \mid \tarray{e_1}{e_2} \mid \tarrget{e_1}{e_2} \mid
        \tarrset{e_1}{e_2}{e_3}\\
v & ::= & \cdots \mid \varr{\ell}{n}\\
T &::=& \cdots \mid \tarrayl{\ell,n}{e_1}{e_2} \mid
\tarrgetl{e_1}{e_2}{\ell[n]} \mid \tarrsetl{e_1}{e_2}{\ell[n]}{e_3}
\end{eqnarray*}
where $\tarray{e_1}{e_2}$ creates an array of length $e_1$ whose elements are
initialised to $e_2$, while $\tarrget{e_1}{e_2}$ gets element $e_2$ from array
$e_1$ and $\tarrset{e_1}{e_2}{e_3}$ assigns $e_3$ to $e_1[e_2]$.  Array values
$\varr{\ell}{n}$ consist of a store location $\ell$ and length $n$.
Furthermore, we extend stores to map locations to either ordinary values $v$
or arrays $[v_1,\ldots,v_n]$. Figure~\ref{fig:arrays-semantics} sketches the
semantics of arrays.  We omit routine additional rules for reading the length
of an array.  Aside from the fact that they record traces, the evaluation
rules are otherwise straightforward.

\begin{figure}
\begin{ruleblock}{T ::\rho,\mu,M \eval \mu',R}
  \inferrule*{
    \rho,e_1 \eval n\\
    \rho,e_2 \eval v
}
{
  \derivesShade{\tarrayl{\ell,n}{e_1}{e_2}}{\rho,\mu,\tarray{e_1}{e_2} \eval \mu[\ell \mapsto [v,\ldots,v]], \rret~{\varr{\ell}{n}}}
}
\and
\inferrule*{
  \rho,e_1 \eval \varr{\ell}{n}\\
\rho,e_2 \eval i\\
0 \leq i < n
}
{
\derivesShade{\tarrgetl{e_1}{e_2}{\ell[i]}}{\rho,\mu, \tarrget{e_1}{e_2} \eval \mu,\rret~{\mu[\ell[i]]}}
}
\and
\inferrule*{
  \rho,e_1 \eval \varr{\ell}{n}\\
\rho,e_2 \eval i\\
\rho,e_3 \eval v\\
0 \leq i < n
}
{
\derivesShade{\tarrsetl{e_1}{e_2}{\ell[i]}{e_3}}{\rho,\mu, \tarrset{e_1}{e_2}{e_3} \eval \mu[\ell[i] = v], \rret~{()}}
}
\end{ruleblock}
\caption{Traced evaluation for array constructs}\label{fig:arrays-semantics}
  \begin{ruleblock}{\rho,\mu,M,T \fwd \mu',R}
  \inferrule*{
    \rho,e_1 \fwd n\\
    \rho,e_2 \fwd v
}
{
\rho,\mu,\tarray{e_1}{e_2},\tarrayl{\ell,n}{e_1}{e_2} \fwd \mu[\ell
\mapsto [v,\ldots,v]], \rret \varr{\ell}{n}
}
\and
  \inferrule*{
    \rho,e_1 \fwd \Hole
}
{
\rho,\mu,\tarray{e_1}{e_2},\tarrayl{\ell,n}{e_1}{e_2} \fwd \mu[\ell \mapsto [\Hole,\ldots,\Hole]], \rret~{\Hole}
}
\and
\inferrule*{
  \rho,e_1 \fwd \varr{\ell}{n}\\
\rho,e_2 \fwd i
}
{
\rho,\mu, \tarrget{e_1}{e_2}, \tarrgetl{e_1}{e_2}{\ell[i]} \fwd \mu,\rret{\mu[\ell[i]]}
}
\and
\inferrule*{
  \rho,e_1 \fwd \Hole\\
\text{or}\\
\rho,e_2 \fwd \Hole
}
{
\rho,\mu, \tarrget{e_1}{e_2}, \tarrgetl{e_1}{e_2}{\ell[i]} \fwd \mu,\rret{\Hole}
}
\and
\inferrule*{
  \rho,e_1 \fwd \varr{\ell}{n}\\
\rho,e_2 \fwd i \\
\rho,e_3 \fwd v
}
{
\rho,\mu, \tarrset{e_1}{e_2}{e_3},\tarrsetl{e_1}{e_2}{\ell[i]}{e_3} \fwd \mu[\ell[i]\mapsto v], \rret{()}
}
\and
\inferrule*{
  \rho,e_1 \fwd \Hole\\
\text{or}\\
\rho,e_2 \fwd \Hole
}
{
\rho,\mu, \tarrset{e_1}{e_2}{e_3},\tarrsetl{e_1}{e_2}{\ell[i]}{e_3} \fwd \mu[\ell[i] \mapsto \Hole], \rret{\Hole}
}
 \end{ruleblock}
\caption{Forward slicing for array constructs}\label{fig:arrays-fwd}
\begin{ruleblock}{\mu,R,T \bwd \rho,\mu',M,U}
\inferrule*{
\sqcup_{i=0}^{n-1}\mu(l[i]), e_2 \bwd \rho_2,e_2' \\
n,e_1 \bwd \rho_1,e_1'
}
{
\mu, \rret{v},\tarrayl{\ell,n}{e_1}{e_2} \bwd \rho_1\sqcup\rho_2,\mu[\ell\mapsto \Hole], \tarray{e_1'}{e_2'}, \tarrayl{\ell,n}{e_1'}{e_2'}
}
\and
\inferrule*{
i,e_2 \bwd \rho_2,e_2'\\
\ell,e_1 \bwd \rho_1,e_1'
}
{
\mu,\rret{v},\tarrgetl{e_1}{e_2}{\ell[i]} \bwd \rho_1\sqcup \rho_2,\mu\sqcup [\ell[i]\mapsto v], \tarrget{e_1'}{e_2'},\tarrgetl{e_1'}{e_2'}{\ell[i]}
}
\and
\inferrule*{
\mu(\ell[i]),e_3 \bwd \rho_3, e_3'\\
i,e_2 \bwd \rho_2,e_2'\\
l,e_1 \bwd \rho_1,e_1'
}
{
\mu,\rret{v},\tarrsetl{e_1}{e_2}{\ell[i]}{e_3} \bwd \rho_1\sqcup \rho_2 \sqcup \rho_3,\mu[\ell[i]\mapsto \Hole], \tarrset{e_1'}{e_2'}{e_3'},\tarrsetl{e_1'}{e_2'}{\ell[i]}{e_3'}
}
\end{ruleblock}
\caption{Backward slicing for array constructs}\label{fig:arrays-bwd}
\end{figure}

Traces for array creation are annotated with the location and length of the
array, while dereference and update operations are annotated with the array
location and affected index. We extend the function $\writes(T)$ as follows:
\begin{eqnarray*}
\writes(\tarrayl{\ell,n}{e_1}{e_2}) &=& \{\ell[0],\ldots,\ell[n-1]\}\\
\writes(\tarrgetl{e_1}{e_2}{\ell[i]}) &=& \{\ell[i]\}\\
\writes(\tarrsetl{e_1}{e_2}{\ell[i]}{e_2}) &=& \{\ell[i]\}
\end{eqnarray*}
We simply define $\res(T)$ as $\mathtt{val}$ for array traces $T$.
(Alternatively, we could instead adjust the semantics of arrays so
that exceptions are raised in the event of attempt to create an array
of negative length or read or write to an out-of-bounds index.  In
that case we would need to annotate traces to reflect these
possibilities, but we omit this added complication.)

Figure~\ref{fig:arrays-fwd} shows the forward slicing rules for
array constructs, which are similar to those for references.  The main
differences are that in the rules for dereferencing and updating, we
require both the array and index parameter to be defined in order to
return a value, and return $\Hole$ if either argument is $\Hole$.  In
that case, we also approximate the effect of the read or write on the
store.

Figure~\ref{fig:arrays-bwd} shows the backward slicing rules for arrays. These
are again similar to those for references.  In the case for array creation, we
use the location and length of the created array to compute the join of all
demanded parts of the initialisation expression $e_2$, and we also require the
length $n$ be recomputed from $e_1$.  In the rule for backward slicing for
array dereferences, we slice $e_2$ with respect to $i$ and $e_1 $ with respect
to $\ell$, where the annotation $\ell[i]$ records the array location and
index; we also place demand $v$ on the $n$th element of the array at $\ell$ in
the store.  Finally, in the backward rule for array update, using the recorded
location and index $\ell[i]$, we slice $e_3$ with respect to the current
demand on $\ell[i]$, and slice the index and array subexpressions as before.
Finally we erase the $i$th element of the array at $\ell$ since its value
before the update is no longer relevant until some earlier computation reads
it.

Sequential composition and while-loops are definable in iTML in the
usual way:
\begin{eqnarray*}
M_1 ; M_2 &\iff& \tlet{\_}{ M_1}{ M_2}\\
\twhile{e}{M} &\iff& (\tfun{loop}{\_}{\tif{e}{(M;loop~\tunit)}{\tunit}})~\tunit
\end{eqnarray*}
Our implementation supports these constructs directly, rather than via
desugaring, so that slicing results in comprehensible slices in terms
of these constructs.  As a simple example illustrating all of the
above features, consider the program in
Figure~\ref{fig:array-example}(a), which creates an array and adds up
the numbers in even positions, and writes the partial sums to the odd
positions.  Slices are shown with respect to the final values of $!s$, $!i$,
and $x[3]$ in Figure~\ref{fig:array-example}(b--d) respectively.

\begin{figure}
  \footnotesize
\begin{tabular}[t]{p{3cm}p{3cm}p{3cm}p{3cm}}
(a) & (b) & (c) & (d)
\smallskip\\
\begin{Verbatim}[commandchars=\\\{\}]
let x = [|0,1,2,3|] in 
let i = ref 0 in 
let s = ref 0 in 
while (!i < 4) do ( 
  s := !s + x[!i]; 
  x[!i+1] <- s;
  i := !i + 2 
)
\end{Verbatim}
&
\begin{Verbatim}[commandchars=\\\{\}]
let x = [|0,\(\slice{1}\),2,\(\slice{3}\)|] in 
let i = ref 0 in 
let s = ref 0 in 
while (!i < 4) do ( 
  s := !s + x[!i]; 
  \(\slice{x[!i+1] <- s}\);
  i := !i + 2 
) 
\end{Verbatim}
&
\begin{Verbatim}[commandchars=\\\{\}]
let x = \(\slice{[|0,1,2,3|]}\) in 
let i = ref 0 in 
let s = \(\slice{ref 0}\) in 
while (!i < 4) do ( 
  \(\slice{s := !s + x[!i])}\)
  \(\slice{x[!i+1] <- s}\);
  i := !i + 2 
)
\end{Verbatim}
&
\begin{Verbatim}[commandchars=\\\{\}]
let x = [|0,\(\slice{1}\),2,\(\slice{3}\)|] in 
let i = ref 0 in 
let s = \slice{ref 0} in 
while (!i < 4) do ( 
  s := !s + x[!i];
  x[!i+1] <- s;
  i := !i + 2 
)
\end{Verbatim}
\end{tabular}
  \caption{Example of slicing using arrays and while-loops
  (a) complete program, (b) slice with respect to $!s = 2$, (c) slice
  with respect to $!i = 4$, (d) slice with respect to $x[3] = 2$}
  \label{fig:array-example}
\end{figure}

\section{Implementation}
\label{sec:implementation}

To validate the ideas presented in the earlier sections we created an
implementation\footnote{\url{https://github.com/jstolarek/slicer}} in Haskell
(GHC 8.0.1) that allows us to run, trace, and backward slice iTML programs,
along with a read-eval-print loop that allows interactive use of these
features.

The calculus introduced in Section~\ref{sec:calculus} is designed to
reduce the number of necessary semantic rules and at the same time
maintain the full expressive power of an ML-like language.  The actual
iTML language in our implementation does not distinguish between
\emph{expressions} and \emph{computations}, which means that
side-effecting, exception-raising computations can occur anywhere.
Indeed, even constructs like nested exceptions
($\traise{(\traise{e}))}$ are permitted.  iTML also contains integer,
double, string and boolean types, arithmetic and logical operators,
pair types with projections, arrays, \texttt{if} conditionals,
sequencing, and loops.

To implement backward slicing algorithm we had to generalise the
slicing rules from
Figures~\ref{fig:slicing-expression},~\ref{fig:slicing-bwd-coarse},
and~\ref{fig:arrays-bwd} to the full iTML language.  As expected, this
causes a blowup in the number of rules, from a total of twenty-five
rules to over seventy cases in the actual code.  Eliminating the
distinction between expressions and computations also leads to the
structure of traces being significantly different from the one shown
in Fig.~\ref{sec:calculus}.  In our core calculus we have two
different trace forms for \texttt{let} expressions to distinguish
whether a \texttt{let}-bound expression raised an exception or not,
and similarly for \texttt{try}-\texttt{with} blocks.  Given the much
richer structure of expressions in the actual implementation, an
approach of having several trace forms for each expression form would
be impractical.  So when a subexpression of a trace raises an
exception we simply denote all remaining sub-traces as $\Hole$.  So,
for example, we represent $\tletFail{T}$ as
$\mathtt{let}({T},x.{\Hole})$.

To evaluate the practical usefulness of our development we decided to
implement a non-trivial algorithm that relies on side-effects and may
potentially raise exceptions.  We picked the Gaussian elimination method for
solving systems of linear equations.  Our implementation is naive: it does not
perform pivoting nor does it try to detect situations where a system has
infinitely many solutions or no solutions at all.  This means that for such
systems our program will attempt a division by zero, thus raising an
exception. Fig.~\ref{fig:gauss}(a) shows a matrix of coefficients in a 4-by-4
system of linear equations.  The first iteration leads to zeroing of elements
below the diagonal in the first column, but it also leads to zeroing of an
element on the diagonal in the second column (Fig.~\ref{fig:gauss}(b), boxed).
In the second iteration we immediately attempt to divide $2\frac{2}{3}$ by
$0$, which leads to an exception being raised.  If we now slice the program
with respect to the exception value, our implementation will identify elements
of the entry matrix that were relevant in raising the exception -- see
Fig.~\ref{fig:gauss}(c), where the boxed element is where the exception is
raised.  Our implementation also identifies which expressions in the program
were relevant to raising an exception.  But since there is only one place
where a division occurs in the code it is pretty obvious from the start where
the division by zero must have taken place. The program slice (shown in full
in \appref{impl-example}) does exclude some code that was not relevant to the
exception, but analysing a trace slice might be much more enlightening here.

Note that system in Fig.~\ref{fig:gauss}(a) has exactly one solution and if we
swap the 2nd and 3rd row our implementation will find it.  To test our
implementation in a higher-order setting we mapped the solving function over a
list of matrices that first contained a solvable modification of the system in
Fig.~\ref{fig:gauss}(a) and then the original version that leads to division by
zero.  Our implementation correctly identifies the first matrix as irrelevant to
the exception result and marks the same elements of second array as the ones shown in
Fig.~\ref{fig:gauss}(c).

\begin{figure}
\small
\begin{tabular}[t]{p{4.5cm}p{5cm}p{4cm}}
(a) & (b) & (c)
\smallskip\\
\(\displaystyle
 \left [
 \begin{array}{rrrr|r}
  3  & -1 &  2 & -1 & -13 \\
  3  & -1 &  1 &  1 &   1 \\
  1  &  2 & -1 &  2 &  21 \\
 -1  &  1 & -2 & -3 &  -5 \end{array} \right ]
\)
&
\(\displaystyle
\left [
 \begin{array}{rrrr|r}
  3  & -1 &  2 & -1 & -13 \\
  0  &  \fbox{0} & -1 &  2 & 14 \\[5pt]
  0  &  2\frac{2}{3} & -1\frac{2}{3} &  2\frac{2}{3} & 25\frac{1}{3} \\[5pt]
  0  &  \frac{2}{3} & -1\frac{1}{3} & -3\frac{1}{3} & -9\frac{1}{3} \end{array} \right ]
\)
&
\(\displaystyle
\left [
 \begin{array}{rrrr|r}
  3  & -1 & \Hole & \Hole & \Hole \\
  3  & -1 & \Hole & \Hole & \Hole \\
  1  &  \fbox{2} & \Hole & \Hole & \Hole \\
  \Hole & \Hole & \Hole & \Hole & \Hole \end{array} \right ]
\)
\end{tabular}
\caption{Gaussian elimination example. (a) initial matrix; (b) after one
  iteration a $0$ appears on the diagonal; (c) our program identifies relevant
  elements of original array, $2$ in a box indicating place where division
  by zero occurred.}
\label{fig:gauss}
\end{figure}

Our main focus has been on developing an intuitively plausible forward
slicing semantics and matching backward semantics that provides useful
information in the presence of side-effects, and our implementation
has been helpful for establishing the usefulness of this approach.
Though achieving high performance has not been our focus, it is also
an important concern, so we have conducted preliminary
investigations of the performance of our approach, for example by
tracing and slicing computations that create lists or arrays of
various lengths.  Our initial approach to backward
slicing recomputes $\writes(T)$ whenever the \rulename{B-Slice$\Hole$}
rule is attempted, and is observed to be quadratic in some cases.
Understanding the performance of the Haskell implementation is
nontrivial and we plan to investigate more efficient techniques in
future work.

\section{Related work}
\label{sec:related}

Galois connections are widely used in (static) program analysis in the
context of \emph{abstract
  interpretation}~\citep{cousot77popl,darais16icfp}.  In that setting,
one lattice might be the (infinite) set of sets of possible run-time
behaviours of a program and another might be the (finite) set of
abstractions computed by a static analysis.  Abstract interpretation
has also recently been related to gradual typing~\citep{garcia16popl},
a technique for mixing static and dynamic type systems.  Here one
lattice is the set of sets of (concrete) types and another is the set
of gradual types.  However, both abstract interpretation and gradual
typing are aimed at static analysis or typechecking of programs,
whereas we consider dynamic analysis via Galois connections between
lattices of partial inputs and partial outputs of a program run.  On
the other hand, it is an intriguing question whether the forward
slicing semantics can be derived from ordinary evaluation using Galois
connections between partial objects and sets of complete objects
(analogously to the AGT approach of~\citet{garcia16popl} but at the
expression/value level).

The application of Galois connections to program slicing for pure higher-order
programs with pairs, sums and recursive datatypes was introduced by
\citeauthor{perera12icfp}~\citeyear{perera12icfp,perera13}. Subsequent work
investigated applications of related slicing techniques to security and
provenance analysis~\citep{acar13jcs} and explaining database
queries~\citep{cheney14ppdp}, although these papers did not employ Galois
connections, opting instead for semantic notions of dependence (based on
replaying traces) for which minimal slicing is undecidable.
\citeauthor{perera16concur} extended the Galois slicing approach to the
$\pi$-calculus~\citeyear{perera16concur}.  Our work draws on their insight
that trace information needs to be taken into account in the definitions of
forward and backward slicing, but we consider a core language iTML for an
ML-like language with imperative features, rather than  the $\pi$-calculus. In
principle it may be possible to translate iTML to the $\pi$-calculus, and use
\citeauthor{perera16concur}'s \citeyear{perera16concur} slicing technique on
the results, but it is unclear how one might translate back to iTML, or
whether the translation would introduce undesirable artefacts.

As noted earlier, there is a large literature on slicing techniques for
imperative and object-oriented languages~\citep{xu05sigsoft}, but to the best
of our knowledge none of this work has been extended to also handle features
common to functional programming languages. Also, to the best of our knowledge
the fact that optimal program slicing techniques are Galois connections has
not been discussed in the slicing literature. \citet{field98ist} present an
approach to slicing for arbitrary sets of rewrite rules, in which forward
slicing and backward slicing enjoy correctness and minimality properties
determined by the rewriting rules.  However, they considered first-order
rewrite systems only, which would not suffice for a higher-order language.

The Galois slicing approach is similar in spirit to several previous papers on
slicing for pure or lazy programs based on recording and analysing redex
trails~\citep{ochoa08hosc,rodrigues07jucs} and semantics-directed execution
monitoring~\citep{kishon95jfp}. \citet{perera12icfp} give a more detailed
comparison with this prior work.  There is also a clear analogy with
\emph{declarative debugging} techniques in logic programming (including
functional-logic programming languages such as Curry and Mercury). For
example, tracing and dependency-tracking techniques have also been used in a
tool for automated debugging in Mercury~\cite{maclarty05aadebug}, in a system
which helps programmers localise bugs by traversing an execution trace in
response to programmer feedback about correct and incorrect results. Work by
\citet{silva06ppdp} on combining algorithmic debugging and program slicing for
pure functional programs could be generalised to automated debugging for
imperative functional programs.

\citet{biswas97phd} did consider slicing for ML programs including references
and effects, but used a semantic notion of program slice for which least
slices are not computable. In his approach, eliding an exception handler can
allow an exception to propagate unhandled or be handled by a different handler
than in the original (unsliced) program. Similarly, eliding an assignment can
expose the previous value of the store location.  This is in contrast with the
Galois connection approach, where slicing is required to be monotone and the
execution of a program slice is always a slice of the original program's
execution. Thus slicing never changes the behaviour of a program, other than
to elide parts in a way consistent with the original execution.

Slicing-like techniques have also been considered recently for explaining type
errors.  Type error explanation and diagnosis in the presence of
Hindley-Milner-style type inference has been studied extensively; we mention a
few closely related approaches. \citet{haack04scp} developed methods for type
error slicing for Standard ML that provide completeness and minimality
guarantees; this suggests that it may be possible to view their approach as a
Galois connection between lattices of programs and type errors.
\citet{seidel16icfp} present an approach for explaining type errors using
\emph{dynamic witnesses}, that is, synthesised input values that illustrate
how the program will go wrong.  Such explanations may be more immediately
useful to novices than conventional type errors, but can grow large; Seidel
\etal suggest that slicing techniques may be useful for providing smaller
explanations.  Our work may apply here, since we can slice programs that are
not well typed.

Bidirectional transformations (such as
\emph{lenses}~\citep{foster10ssgip}) consist of pairs of mappings
between data structures that maintain some notion of consistency among
them; for example, bidirectional transformations are proposed for
synchronising different models of a software system, such as class
diagrams and database schemas. Bidirectional transformations satisfy
round-tripping laws that ensure that changes made to one side of the
transformation are appropriately propagated to the other side. Among
the growing literature on bidirectional programming, the work of
\citet{wang11icfp} seems particularly relevant, since it considers
bidirectional transformations on tree-structured datatypes
(e.g. abstract syntax trees).  In their approach, changes to one side
of the transformation can be propagated efficiently to the other side
by decomposing the tree into a context (which does not change) and a
focused subtree that is changed. It may make sense to view backward
slicing as a special form of bidirectional transformation in which we
only consider deleting subexpressions from the output; the
relationship between Galois slicing and bidirectional transformations
remains to be investigated.

We briefly considered the possibility of lifting slicing for the iTML core
language to ML-like source programs by desugaring, slicing the desugared
programs, and then somehow resugaring the sliced program back to an ML-like
program. \citet{pombrio14pldi,pombrio15icfp} proposed an approach to
resugaring for languages defined compositionally using hygienic macros, so
that evaluation steps in the desugared language can be made meaningful in
terms of the source language.  Their approach establishes equational
properties for round-tripping between desugaring and resugaring, similar to
those encountered in bidirectional transformations or Galois connections.  It
would be interesting to see whether compatible desugaring/resugaring pairs can
be lifted to Galois connections on partial expressions, since we could
potentially then lift slicing to the source language by composing with
desugaring/resugaring.  Other approaches to operational semantics,
such as \citeauthor{chargueraud13esop}'s \citeyear{chargueraud13esop} \emph{pretty big-step semantics}, might also be
worth considering.

Techniques for working with partial programs have also been considered
recently by \citeauthor{omar17popl} in the structured editor system
Hazelnut~\citeyearpar{omar17popl}.  They explore usage of holes as a way to
write programs in incremental steps, while guaranteeing that incomplete
programs are meaningful at each intermediate editing step. Interestingly,
Hazelnut allows holes to take parameters, so that a term that is not
well-typed in the current context can be placed inside a parameterized hole.
It may be fruitful to combine the ideas in our approach to evaluating and slicing partial
programs with Hazelnut's approach to structured editing.

\section{Conclusions}
\label{sec:conclusions}

Despite its long history and extensive exploration in imperative or
object-oriented settings, program slicing is not yet well-understood for
functional languages.  To date, most work on slicing for functional languages
has not considered effects; the main exception is \citet{biswas97phd}, but his
approach is extremely conservative in the presence of effects.  On the other
hand, work on slicing for imperative languages has not considered higher-order
functions, datatypes or other common features of functional languages.

In this paper we generalised the Galois slicing approach, which considers
fine-grained forward and backward slicing techniques as Galois connections
between lattices of partial inputs and outputs, to also allow for traces that
determinise and record information about the effects of computations.  We
defined tracing semantics and forward and backward slicing for an imperative
core language, iTML, and proved that they form a Galois connection. We have
implemented and evaluated our approach on a variety of examples, providing
additional confidence in its usefulness.  Our main contribution is the
definition of forward slicing and matching optimal backward slicing, proofs of
their correctness, and experimental investigation of their qualitative
usefulness.

There are a number of interesting directions for future work. Currently, there
is a gap between slicing for the core language (which we use for proofs) and
the source language, in which we have to handle many additional cases.  It
seems straightforward, albeit labour-intensive, to extend the systems and
proofs; we would prefer to find a more elegant approach that allows us to lift
results about slicing from the core language to the source language through
resugaring and desugaring.  Adapting our approach to a mainstream language may
raise additional issues we have not had to consider in the core language.
Extending our approach to consider other effects, objects, or concurrency
appears to be a considerable challenge. Finally, we have focused on
correctness and expressiveness, so finding efficient techniques for slicing
that can be applied to larger programs is an important next step.

\begin{acks}
  Effort sponsored by the Air Force Office of Scientific Research, Air
  Force Material Command, USAF, under grant number
  FA8655-13-1-3006. The U.S. Government and University of Edinburgh
  are authorised to reproduce and distribute reprints for their
  purposes notwithstanding any copyright notation thereon. Perera was
  also supported by UK EPSRC project EP/K034413/1, and Stolarek and
  Cheney were supported by ERC Consolidator Grant Skye (grant number
  682315).  We are grateful to the anonymous reviewers for
  constructive and encouraging feedback.
\end{acks}

\bibliographystyle{plainnat} 

\def\includeapp{1}
\ifdefined\includeapp
\pagebreak
\appendix
\section{Prefixes of expressions, computations and traces}
\label{app:prefix}

\figref{expression-leq} defines the prefix relation for partial
expressions and partial computations; \figref{trace-leq} defines the prefix
relation for partial traces.

\begin{figure}
  \begin{ruleblock}{e \sqleq e'}
\inferrule*
{\strut}
{\Hole \sqleq e}
\and
\inferrule*
{\strut}
{x \sqleq x}
\and
\inferrule*
{\strut}
{\tunit \sqleq \tunit}
\and
\inferrule*
{e_1 \sqleq e_1' \\
  e_2 \sqleq e_2'}
{\tpair{e_1}{e_2} \sqleq \tpair{e_1'}{e_2'}}
\and
\inferrule*
{e \sqleq e'}
{\tfst{e} \sqleq \tfst{e'}}
\and
\inferrule*
{e \sqleq e'}
{\tsnd{e} \sqleq \tsnd{e'}}
\and
\inferrule*
{e \sqleq e'}
{\tinl{e} \sqleq \tinl{e'}}
\and
\inferrule*
{e \sqleq e'}
{\tinr{e} \sqleq \tinr{e'}}
\and
\inferrule*
{M \sqleq M'}
{\tfun{f}{x}{M} \sqleq \tfun{f}{x}{M'} }
\end{ruleblock}
\begin{ruleblock}{M \sqleq M'}
\inferrule*
{\strut}
{\Hole \sqleq M}
\and
\inferrule*
{e \sqleq e'}
{\tret{e} \sqleq \tret{e'}}
\and
\inferrule*
{e_1 \sqleq e_1' \\
e_2 \sqleq e_2'}
{e_1~e_2 \sqleq e_1'~e_2'}
\and
\inferrule*
{M_1 \sqleq M_1' \\
M_2 \sqleq M_2'}
{\tlet{x}{M_1}{M_2}\sqleq \tlet{x}{M_1'}{M_2'}}
\and
\inferrule*
{e \sqleq e' \\
M_1 \sqleq M_1'\\
M_2 \sqleq M_2'}
{\tcase{e}{\tinlC{x}{M_1}}{\tinrC{y}{M_2}} \sqleq \tcase{e'}{\tinlC{x}{M_1'}}{\tinrC{y}{M_2'}}}
\and
\inferrule*
{e \sqleq e'}
{\traise{e} \sqleq \traise{e'}}
\and
\inferrule*
{M_1 \sqleq M_1' \\
M_2 \sqleq M_2'}
{\ttry{M_1}{x}{M_2}\sqleq \ttry{M_1'}{x}{M_2'}}
\and
\inferrule*
{e \sqleq e'}
{\tref{e} \sqleq \tref{e'}}
\and
\inferrule*
{e_1 \sqleq e_1' \\
e_2 \sqleq e_2'}
{\tassign{e_1}{e_2} \sqleq \tassign{e_1'}{e_2'}}
\and
\inferrule*
{e \sqleq e'}
{\bang e \sqleq \bang e'}
\end{ruleblock}
\caption{Partial expression and partial computation prefix relations}
\label{fig:expression-leq}
\end{figure}

\begin{figure}
  \begin{ruleblock}{T \sqleq T'}
\inferrule*
{\writes(T) = \Ell\\
\res(T) = \kay}
{\thole{\Ell}{\kay} \sqleq T}
\and
\inferrule*
{e \sqleq e'}
{\tret{e} \sqleq \tret{e'}}
\and
\inferrule*
{e_1 \sqleq e_1' \\
e_2 \sqleq e_2'\\
T \sqleq T'}
{\tappBody{e_1}{e_2}{f}{x}{T} \sqleq \tappBody{e_1'}{e_2'}{f}{x}{T'}}
\and
\inferrule*
{T_1 \sqleq T_1' \\
T_2 \sqleq T_2'}
{\tletSucc{x}{T_1}{T_2}\sqleq \tletSucc{x}{T_1'}{T_2'}}
\and
\inferrule*
{T_1 \sqleq T_1'}
{\tletFail{T}\sqleq \tletFail{T'}}
\and
\inferrule*
{e \sqleq e' \\
T \sqleq T'}
{\tcaseInl{e}{x}{T}{y} \sqleq \tcaseInl{e'}{x}{T'}{y}}
\and
\inferrule*
{e \sqleq e' \\
T \sqleq T'}
{\tcaseInr{e}{x}{y}{T} \sqleq \tcaseInr{e'}{x}{y}{T'}}
\and
\inferrule*
{e \sqleq e'}
{\traise{e} \sqleq \traise{e'}}
\and
\inferrule*
{T_1 \sqleq T_1' \\
T_2 \sqleq T_2'}
{\ttryFail{T_1}{x}{T_2}\sqleq \ttryFail{T_1'}{x}{T_2'}}
\and
\inferrule*
{T \sqleq T'}
{\ttrySucc{T}\sqleq \ttrySucc{T'}}
\and
\inferrule*
{e \sqleq e'}
{\trefl{\ell}{e} \sqleq \trefl{\ell}{e'}}
\and
\inferrule*
{e_1 \sqleq e_1' \\
e_2 \sqleq e_2'}
{\tassignl{e_1}{\ell}{e_2} \sqleq \tassignl{e_1'}{\ell}{e_2'}}
\and
\inferrule*
{e \sqleq e'}
{\tderefl{\ell}{e} \sqleq \tderefl{\ell}{ e'}}
\end{ruleblock}
\caption{Partial trace prefix relation}
\label{fig:trace-leq}
\end{figure}

\section{Proofs}

\subsection{Lemmas}
We rely on monotonicity properties for forward and backward slicing,
which are easily checked:

\begin{corollary}[Monotonicity of $\fwdFs{}$]
\label{cor:fwd:monotonicity:expression}
Suppose $\sigma, e \eval v$ and $x \sqleq x' \sqleq
(\sigma, e)$. Then $\fwdFs{\sigma,e}(x) \sqleq \fwdFs{\sigma,e}(x')$.
\end{corollary}

\begin{corollary}[Monotonicity of $\bwdFs{}$]
\label{cor:bwd:monotonicity:expression}
Suppose $\sigma, e \eval v$ and $u \sqleq u' \sqleq v$. Then
$\bwdFs{\sigma,e}(u) \sqleq \bwdFs{\sigma,e}(u')$.
\end{corollary}

\begin{corollary}[Monotonicity of $\fwdFs{T}$]
\label{cor:fwd:monotonicity:computation}
Suppose $\derives{T}{\sigma, \nu, N \eval \sigma', R}$ and $x \sqleq
x' \sqleq (\sigma, \nu, N, T)$. Then $\fwdFs{T}(x) \sqleq \fwdFs{T}(x')$.
\end{corollary}

\begin{corollary}[Monotonicity of $\bwdFs{T}$]
\label{cor:bwd:monotonicity:computation}
Suppose $\derives{T}{\sigma, \nu, M \eval \nu', S}$ and $x \sqleq x' \sqleq (\nu', S)$. Then $\bwdFs{T}(x) \sqleq \bwdFs{T}(x')$.
\end{corollary}

Additionally, we will use the following properties of forward slicing concerning $\res$ and $\writes$.
\begin{lemma}\label{lem:writes-res}
If $T \sqleq T'$, then $\writes(T) = \writes(T')$ and $\res(T) = \res(T')$.
\end{lemma}
\begin{lemma}\label{lem:fwd:trace:res}
If $\rho,\mu,M,T \fwd \nu,R$ and $\res(T) = \kay$, then $R = \kay~v$ for some $v$.
\end{lemma}
\begin{lemma}\label{lem:fwd:trace:writes}
If $\rho,\mu,M,T \fwd \nu,R$ and $\writes(T) = \Ell$, then $\mu \antirestrict \Ell = \nu \antirestrict \Ell$.
\end{lemma}

\subsection{Conventions}

We write $RS$ for the composition of relations $R$ and $S$, so that for
example $X \mathrel{\fwd\sqgeq} Y$ iff there exists $Z$ such that $X \fwd Z
\sqgeq Y$. The proofs for \thmref{galois-connection:computation} involve an
ambient execution, which for convenience is left implicit. When the ambient
execution ensures that $x$ and $y$ are upper-bounded (and thus the join of $x$
and $y$ is defined) we write this assertion as $x \compatible y$. The symbol
$\qedLocalRaw$ indicates that a proof obligation is being discharged, and IH
stands for ``inductive hypothesis'' (or ``induction hypothesis'', if you
prefer). Throughout, we make free use of the fact that all syntactic forms
preserve and reflect $\sqleq$, so that for example $\tinl{v} \sqleq
\tinl{v'}$ if and only if $v \sqleq v'$.

\newcommand*{\derivationWidth}{0.51\textwidth}

\setcounter{equation}{0}
\proofContext{computation-fwd-preserves-meets}
\subsection{\lemref{fwd:meet-preservation:computation}}
\emph{Suppose $\derives{\bar{T}}{\bar{\rho}, \bar{\mu}, \bar{M} \eval \bar{\mu'}, \bar{R}}$ and $Z_1, Z_2 \sqleq (\bar{\rho}, \bar{\mu}, \bar{M}, \bar{T})$. Then $\fwdFs{\BT}(Z_1 \meet Z_2) = \fwdFs{\BT}(Z_1)
\meet \fwdFs{\BT}(Z_2)$.}

In this proof, we fix the following abbreviations:
\begin{align*}
\bar{Z} & = (\bar{\rho}, \bar{\mu}, \bar{M}, \bar{T}) \\
Z_1 & = (\rho_1, \mu_1, M_1, T_1) \\
Z_2 & = (\rho_2, \mu_2, M_2, T_2) \\
Z_\meet & = Z_1 \meet Z_2 = (\rho_\meet, \mu_\meet, M_\meet, T_\meet)
\end{align*}

Hence, we will prove that, given $Z_1, Z_2 \sqleq \bar{Z}$, $\fwdFs{\BT}(Z_\meet) = \fwdFs{\BT}(Z_1) \meet \fwdFs{\BT}(Z_2)$.

We proceed by structural induction on the derivation of $\fwdFs{\BT}(Z_\meet)$. For each rule, $Z_\meet$ has a certain shape that, combined with the hypothesis $Z_1,Z_2 \sqleq \bar{Z}$, allows us to deduce the allowed shape for $Z_1$, $Z_2$, and $\bar{Z}$. For example, if $M_\meet = \tret{e_\sqcap}$, then $M_i = \tret{e_i}$, and furthermore $\bar{M} = \tret{\bar{e}}$, where $e_i \sqleq \bar{e}$; thus we also have $e_\meet = e_1 \meet e_2$. We will use these properties freely in the following proof cases.
\begin{proof}
\small

\end{proof}

\setcounter{equation}{0}
\proofContext{computation-bwd-after-fwd}
\subsection{\thmref{galois-connection:computation}, part (i)}
Induction on the $\fwd$ derivation. 
\begin{proof}
\small
%
\end{proof}

\setcounter{equation}{0}
\proofContext{computation-fwd-after-bwd}
\subsection{\thmref{galois-connection:computation}, part (ii)}
Induction on the $\bwd$ derivation.
\begin{proof}
\small
\begin{flalign}
\intertext{\crossrule}
&
\caseDerivation{\derivationWidth}{
\begin{smathpar}
\inferrule*
{
   \Ell = \writes(T)
   \\
   \mu \antirestrict \Ell = \mu
}
{
   \mu, \kay~\Hole, T 
   \bwd 
   \Hole, \mu, \Hole, \thole{\Ell}{\kay}
}
\end{smathpar}
}
&
\notag
\\
&
(\mu \antirestrict \Ell, \kay~\Hole) = (\mu, \kay~\Hole) \sqgeq (\mu, \kay~\Hole)
&
\text{
   reflexivity
}
\notag
\\
&
\qedLocal
\derivation{\derivationWidth}{
\begin{smathpar}
\inferrule*
{
   \strut
}
{
   \Hole, \mu, \Hole, \thole{\Ell}{\kay}
   \fwd
   \mu \antirestrict \Ell, \kay~\Hole
}
\end{smathpar}
}
&
\notag
\intertext{\crossrule}
&
\caseDerivation{\derivationWidth}{
\begin{smathpar}
\inferrule*
{
   v, e \bwd \rho,  e'
}
{
   \mu, \rret{v}, \tret{e} 
   \bwd
   \rho, \mu, \tret{e'}, \tret{e'}
}
\end{smathpar}
}
&
\notag
\\
&
\rho, e' \fwd v' \sqgeq v
&
\text{
   \thmref{galois-connection:expression}, part (ii)
}
\locallabel{ret-premise}
\\
&
\qedLocal
\derivation{\derivationWidth}{
\begin{smathpar}
\inferrule*
{
   \rho, e' \fwd v'
}
{
   \rho, \mu, \tret{e'}, \tret{e'}
   \fwd 
   \mu, \rret{v'}
}
\end{smathpar}
}
&
\text{
   (\localref{ret-premise}) 
}
\notag
\intertext{\crossrule}
&
\caseDerivation{\derivationWidth}{
\begin{smathpar}
\inferrule*
{
   \mu, R, T_2 \bwd \extend{\rho_2}{x}{v}, \mu', M_2, U_2
   \\
   \mu', \rret{v}, T_1 \bwd \rho_1, \mu^\twoPrime, M_1, U_1
}
{
   \mu, R, \tletSucc{x}{T_1}{T_2} 
   \bwd
   \rho_1 \join \rho_2, \mu^\twoPrime, \tlet{x}{M_1}{M_2}, \tletSucc{x}{U_1}{U_2}
}
\end{smathpar}
}
&
\notag
\\
&
\rho_1, \mu^\twoPrime, M_1, U_1 \mathrel{\fwd\sqgeq} \mu', \rret{v}
&
\text{
   IH
}
\notag
\\
&
\rho_1 \join \rho_2, \mu^\twoPrime, M_1, U_1 
\fwd 
\mu^\dagger, \rret{v'} \sqgeq \mu', \rret{v}
&
\text{
   $\rho_1 \compatible \rho_2$;
   \corref{fwd:monotonicity:computation}
}
\locallabel{letSucc-premise-one}
\\
&
\extend{\rho_2}{x}{v}, \mu', M_2, U_2  \mathrel{\fwd\sqgeq} \mu, R
&
\text{
   IH
}
\notag
\\
&
\extend{(\rho_1 \join \rho_2)}{x}{v'}, \mu^\dagger, M_2, U_2 
\fwd 
\mu^\ddagger, R' \sqgeq \mu, R
&
\text{
   $\rho_1 \compatible \rho_2$;
   $(v', \mu^\dagger) \sqgeq (v, \mu')$;
   \corref{fwd:monotonicity:computation}
}
\locallabel{letSucc-premise-two}
\\
&
\qedLocal
\derivation{\derivationWidth}{
\begin{smathpar}
\inferrule*
{
   \rho_1 \join \rho_2, \mu^\twoPrime, M_1, U_1 
   \fwd 
   \mu^\dagger, \rret{v'}
   \\
   \extend{(\rho_1 \join \rho_2)}{x}{v'}, \mu^\dagger, M_2, U_2 
   \fwd 
   \mu^\ddagger, R'
}
{
   \rho_1 \join \rho_2, \mu^\twoPrime, \tlet{x}{M_1}{M_2}, \tletSucc{x}{U_1}{U_2}
   \fwd
   \mu^\ddagger, R'
}
\end{smathpar}
}
&
\text{
   (\localref{letSucc-premise-one}, 
    \localref{letSucc-premise-two}) 
}
\notag
\intertext{\crossrule}
&
\caseDerivation{\derivationWidth}{
\begin{smathpar}
\inferrule*
{
   \mu, \rraise{v}, T_1 \bwd \rho, \mu', M_1, U_1
}
{
   \mu, \rraise{v}, \tletFail{T_1}
   \bwd
   \rho, \mu', \tlet{x}{M_1}{\Hole}, \tletFail{U_1}
}
\end{smathpar}
}
&
\notag
\\
&
\rho, \mu', M_1, U_1 \fwd \mu^\twoPrime, \rraise{v'} \sqgeq \mu', \rraise{v}
&
\text{
   IH
}
\locallabel{letFail-premise}
\\
&
\qedLocal
\derivation{\derivationWidth}{
\begin{smathpar}
\inferrule*{
   \rho, \mu', M_1, U_1 \fwd \mu^\twoPrime, \rraise{v'}
}
{
   \rho, \mu', \tlet{x}{M_1}{\Hole}, \tletFail{U_1}
   \fwd
   \mu^\twoPrime, \rraise{v'}
}
\end{smathpar}
}
&
\text{
   (\localref{letFail-premise}) 
}
\notag
\intertext{\crossrule}
&
\caseDerivation{\derivationWidth}{
\begin{smathpar}
\inferrule*
{
   \mu, R, T \bwd \extend{\rho}{x}{v}, \mu', M_1, U
   \\
   \tinl{v}, e \bwd \rho', e'
}
{
   \mu, R, \tcaseInl{e}{x}{T}{y}
   \bwd 
   \rho \join \rho', \mu', \tcase{e'}{\tinlC{x}{M_1}}{\tinrC{y}{\Hole}}, \tcaseInl{e'}{x}{U}{y}
}
\end{smathpar}
}
&
\notag
\\
&
\rho', e' \mathrel{\fwd\sqgeq} \tinl{v}
&
\text{
   \thmref{galois-connection:expression}, part (ii)
}
\notag
\\
&
\rho \join \rho', e' \fwd \tinl{v'} \sqgeq \tinl{v}
&
\text{
   $\rho \compatible \rho'$;
   \corref{fwd:monotonicity:expression}
}
\locallabel{caseInl-premise-one}
\\
&
\extend{\rho}{x}{v}, \mu', M_1, U \mathrel{\fwd\sqgeq} \mu, R
&
\text{
   IH
}
\notag
\\
&
\extend{(\rho \join \rho')}{x}{v'}, \mu', M_1, U \fwd \mu^\twoPrime, R' \sqgeq \mu, R
&
\text{
   $\rho \compatible \rho'$;
   $v' \sqgeq v$;
   \corref{fwd:monotonicity:computation}
}
\locallabel{caseInl-premise-two}
\\
&
\qedLocal
\derivation{\derivationWidth}{
\begin{smathpar}
\inferrule*
{
   \rho \join \rho', e' \fwd \tinl{v'}
   \\
   \extend{(\rho \join \rho')}{x}{v'}, \mu', M_1, U \fwd \mu^\twoPrime, R'
}
{
   \rho \join \rho', \mu', \tcase{e'}{\tinlC{x}{M_1}}{\tinrC{y}{\Hole}}, \tcaseInl{e'}{x}{U}{y}
   \fwd 
   \mu^\twoPrime, R'
}  
\end{smathpar}
}
&
\text{
   (\localref{caseInl-premise-one}, 
    \localref{caseInl-premise-two}) 
}
\notag
\intertext{\crossrule}
&
\caseDerivation{\derivationWidth}{
\begin{smathpar}
\inferrule*
{
   \mu, R, T \bwd \extend{\rho}{y}{v}, \mu', M_2, U
   \\
   \tinr{v}, e \bwd \rho', e'
}
{
   \mu, R, \tcaseInr{e}{x}{y}{T} 
   \bwd 
   \rho \join \rho', \mu', \tcase{e'}{\tinlC{x}{\Hole}}{\tinrC{y}{M_2}}, \tcaseInr{e'}{x}{y}{U}
}
\end{smathpar}
}
&
\notag
\\
&
\rho', e' \mathrel{\fwd\sqgeq} \tinr{v}
&
\text{
   \thmref{galois-connection:expression}, part (ii)
}
\notag
\\
&
\rho \join \rho', e' \fwd \tinr{v'} \sqgeq \tinr{v}
&
\text{
   $\rho \compatible \rho'$;
   \corref{fwd:monotonicity:expression}
}
\locallabel{caseinr-premise-one}
\\
&
\extend{\rho}{y}{v}, \mu', M_2, U \mathrel{\fwd\sqgeq} \mu, R
&
\text{
   IH
}
\notag
\\
&
\extend{(\rho \join \rho')}{y}{v'}, \mu', M_2, U \fwd \mu^\twoPrime, R' \sqgeq \mu, R
&
\text{
   $\rho \compatible \rho'$;
   $v' \sqgeq v$;
   \corref{fwd:monotonicity:computation}
}
\locallabel{caseinr-premise-two}
\\
&
\qedLocal
\derivation{\derivationWidth}{
\begin{smathpar}
\inferrule*
{
   \rho \join \rho', e' \fwd \tinr{v'}
   \\
   \extend{(\rho \join \rho')}{y}{v'}, \mu', M_2, U \fwd \mu^\twoPrime, R'
}
{
   \rho \join \rho', \mu', \tcase{e'}{\tinrC{x}{\Hole}}{\tinrC{y}{M_2}}, \tcaseInr{e'}{x}{y}{U}
   \fwd 
   \mu^\twoPrime, R'
}  
\end{smathpar}
}
&
\text{
   (\localref{caseinr-premise-one}, 
    \localref{caseinr-premise-two}) 
}
\notag
\intertext{\crossrule}
&
\caseDerivation{\derivationWidth}{
\begin{smathpar}
\inferrule*
{
   \mu, R, T 
   \bwd 
   \extend{\extend{\rho}{f}{v_1}}{x}{v_2}, \mu', M, U
   \\
   v_2, e_2 \bwd \rho_2, e_2'
   \\
   v_1 \join \tclosure{\rho}{f}{x}{M}, e_1 \bwd \rho_1, e_1'
}
{
   \mu, R, \tappBody{e_1}{e_2}{f}{x}{T} 
   \bwd 
   \rho_1 \join \rho_2, \mu', \tapp{e_1'}{e_2'}, \tappBody{e_1'}{e_2'}{f}{x}{U} 
}
\end{smathpar}
}
&
\notag
\\
&
\rho_1, e_1' 
\mathrel{\fwd\sqgeq}
v_1 \join \tclosure{\rho}{f}{x}{M} 
&
\text{
   \thmref{galois-connection:expression}, part (ii)
}
\notag
\\
&
\rho_1 \join \rho_2, e_1' \fwd v_1' 
\sqgeq v_1 \join \tclosure{\rho}{f}{x}{M} 
&
\text{
   $\rho_1 \compatible \rho_2$;
   \corref{fwd:monotonicity:expression}
}
\locallabel{app-premise-one}
\\
&
\rho_2, e_2' 
\mathrel{\fwd\sqgeq}
v_2  
&
\text{
   \thmref{galois-connection:expression}, part (ii)
}
\notag
\\
&
\rho_1 \join \rho_2, e_2' \fwd v_2'
\sqgeq v_2  
&
\text{
   $\rho_1 \compatible \rho_2$;
   \corref{fwd:monotonicity:expression}
}
\locallabel{app-premise-two}
\\
&
v_1' = \tclosure{\rho'}{f}{x}{M'}\text{ where }
(\rho', M') \sqgeq (\rho, M)
&
\text{
   (\localref{app-premise-one})
}
\notag
\\
&
\extend{\extend{\rho}{f}{v_1}}{x}{v_2}, \mu', M
, U
\mathrel{\fwd\sqgeq}
\mu, R
&
\text{
   IH
}
\notag
\\
&
\extend{\extend{\rho'}{f}{v_1'}}{x}{v_2'}, \mu', M', U
\fwd
\mu^\twoPrime, R' \sqgeq \mu, R
&
\text{
   $(\rho', v_1', v_2', M') \sqgeq (\rho, v_1, v_2, M)$;
   \corref{fwd:monotonicity:computation}
}
\locallabel{app-premise-three}
\\
&
\qedLocal
\derivation{\derivationWidth}{
\begin{smathpar}
\inferrule*
{
   \rho_1 \join \rho_2, e_1' \fwd v_1' 
   \\
   v_1' = \tclosure{\rho'}{f}{x}{M'}
    \\
   \rho_1 \join \rho_2, e_2' \fwd v_2'
   \\
   \extend{\extend{\rho'}{f}{v_1'}}{x}{v_2'}, \mu', M', U
   \fwd
   \mu^\twoPrime, R'
}
{
   \rho_1 \join \rho_2, \mu', \tapp{e_1'}{e_2'}, \tappBody{e_1'}{e_2'}{f}{x}{U}
   \fwd 
   \mu^\twoPrime, R'
}
\end{smathpar}
}
&
\text{
   (\localref{app-premise-one}, 
    \localref{app-premise-two}, 
    \localref{app-premise-three})
}
\notag
\intertext{\crossrule}
&
\caseDerivation{\derivationWidth}{
\begin{smathpar}
\inferrule*
{
   v, e \bwd \rho,  e'
}
{
   \mu, \rraise{v}, \traise{e} 
   \bwd
   \rho, \mu, \traise{e'}, \traise{e'}
}
\end{smathpar}
}
&
\notag
\\
&
\rho, e' \fwd v' \sqgeq v
&
\text{
   \thmref{galois-connection:expression}, part (ii)
}
\locallabel{raise-premise}
\\
&
\qedLocal
\derivation{\derivationWidth}{
\begin{smathpar}
\inferrule*
{
   \rho, e' \fwd v'
}
{
   \rho, \mu, \traise{e'}, \traise{e'}
   \fwd 
   \mu, \rraise{v'}
}
\end{smathpar}
}
&
\text{
   (\localref{raise-premise}) 
}
\notag
\intertext{\crossrule}
&
\caseDerivation{\derivationWidth}{
\begin{smathpar}
\inferrule*
{
   \mu, R, T_2 \bwd \extend{\rho_2}{x}{v}, \mu', M_2, U_2
   \\
   \mu', \rraise{v}, T_1 \bwd \rho_1, \mu^\twoPrime, M_1, U_1
}
{
   \mu, R, \ttryFail{T_1}{x}{T_2} 
   \bwd
   \rho_1 \join \rho_2, \mu^\twoPrime, \ttry{M_1}{x}{M_2}, \ttryFail{U_1}{x}{U_2}
}
\end{smathpar}
}
&
\notag
\\
&
\rho_1, \mu^\twoPrime, M_1, U_1 \mathrel{\fwd\sqgeq} \mu', \rraise{v}
&
\text{
   IH
}
\notag
\\
&
\rho_1 \join \rho_2, \mu^\twoPrime, M_1, U_1 
\fwd 
\mu^\dagger, \rraise{v'} \sqgeq \mu', \rraise{v}
&
\text{
   $\rho_1 \compatible \rho_2$;
   \corref{fwd:monotonicity:computation}
}
\locallabel{tryFail-premise-one}
\\
&
\extend{\rho_2}{x}{v}, \mu', M_2, U_2  \mathrel{\fwd\sqgeq} \mu, R
&
\text{
   IH
}
\notag
\\
&
\extend{(\rho_1 \join \rho_2)}{x}{v'}, \mu^\dagger, M_2, U_2 
\fwd 
\mu^\ddagger, R' \sqgeq \mu, R
&
\text{
   $\rho_1 \compatible \rho_2$;
   $(v', \mu^\dagger) \sqgeq (v, \mu')$;
   \corref{fwd:monotonicity:computation}
}
\locallabel{tryFail-premise-two}
\\
&
\qedLocal
\derivation{\derivationWidth}{
\begin{smathpar}
\inferrule*
{
   \rho_1 \join \rho_2, \mu^\twoPrime, M_1, U_1 
   \fwd 
   \mu^\dagger, \rraise{v'}
   \\
   \extend{(\rho_1 \join \rho_2)}{x}{v'}, \mu^\dagger, M_2, U_2 
   \fwd 
   \mu^\ddagger, R'
}
{
   \rho_1 \join \rho_2, \mu^\twoPrime, \ttry{M_1}{x}{M_2}, \ttryFail{U_1}{x}{U_2}
   \fwd
   \mu^\ddagger, R'
}
\end{smathpar}
}
&
\text{
   (\localref{tryFail-premise-one}, 
    \localref{tryFail-premise-two}) 
}
\notag
\intertext{\crossrule}
&
\caseDerivation{\derivationWidth}{
\begin{smathpar}
\inferrule*
{
   \mu, \rret{v}, T_1 \bwd \rho, \mu', M_1, U_1
}
{
   \mu, \rret{v}, \ttrySucc{T_1}
   \bwd
   \rho, \mu', \ttry{M_1}{x}{\Hole}, \ttrySucc{U_1}
}
\end{smathpar}
}
&
\notag
\\
&
\rho, \mu', M_1, U_1 \fwd \mu^\twoPrime, \rret{v'} \sqgeq \mu', \rret{v}
&
\text{
   IH
}
\locallabel{trySucc-premise}
\\
&
\qedLocal
\derivation{\derivationWidth}{
\begin{smathpar}
\inferrule*{
   \rho, \mu', M_1, U_1 \fwd \mu^\twoPrime, \rret{v'}
}
{
   \rho, \mu', \ttry{M_1}{x}{\Hole}, \ttrySucc{U_1}
   \fwd
   \mu^\twoPrime, \rret{v'}
}
\end{smathpar}
}
&
\text{
   (\localref{trySucc-premise}) 
}
\notag
\intertext{\crossrule}
&
\caseDerivation{\derivationWidth}{
\begin{smathpar}
\inferrule*
{
   \mu(\ell), e \bwd \rho, e'
}
{
   \mu, \rret{v}, \trefl{\ell}{e} 
   \bwd 
   \rho, \update{\mu}{\ell}{\Hole}, \tref{e'}, \trefl{\ell}{e'}
}
\end{smathpar}
}
&
\notag
\\
&
\rho, e' \fwd v' \sqgeq \mu(\ell)
&
\text{
   \thmref{galois-connection:expression}, part (ii)
}
\locallabel{ref-premise}
\\
&
\rret{v} \sqleq \rret{\ell}
&
\text{
   inversion on $\eval$
}
\notag
\\
&
\qedLocal
(\update{\update{\mu}{\ell}{\Hole}}{\ell}{v'}, \ell) \sqgeq (\mu, v)
&
\text{
   $v' \sqgeq \mu(\ell)$
}
\notag
\\
&
\qedLocal
\derivation{\derivationWidth}{
\begin{smathpar}
\inferrule*
{
   \rho, e' \fwd v' 
}
{
   \rho, \update{\mu}{\ell}{\Hole}, \tref{e'}, \trefl{\ell}{e'}
   \fwd
   \update{\update{\mu}{\ell}{\Hole}}{\ell}{v'}, \rret{\ell}
}
\end{smathpar}
}
&
\text{
   (\localref{ref-premise}) 
}
\notag
\intertext{\crossrule}
&
\caseDerivation{\derivationWidth}{
\begin{smathpar}
\inferrule*
{
   \mu(\ell) \neq \Hole
   \\
   \mu(\ell), e_2 \bwd \rho_2, e_2'
   \\
   \ell, e_1 \bwd \rho_1, e_1'
}
{
   \mu, \rret{v}, \tassignl{e_1}{\ell}{e_2} 
   \bwd 
   \rho_1 \join \rho_2, 
   \update{\mu}{\ell}{\Hole}, \tassign{e_1'}{e_2'}, \tassignl{e_1'}{\ell}{e_2'}
}
\end{smathpar}
}
&
\notag
\\
&
\rho_1, e_1' \mathrel{\fwd\sqgeq} \ell
&
\text{
   \thmref{galois-connection:expression}, part (ii)
}
\notag
\\
&
\rho_1 \join \rho_2, e_1' \fwd \ell \sqgeq \ell
&
\text{
   $\rho_1 \compatible \rho_2$;
   \corref{fwd:monotonicity:expression}
}
\locallabel{assign-premise-one}
\\
&
\rho_2, e_2' \mathrel{\fwd\sqgeq} \mu(\ell)
&
\text{
   \thmref{galois-connection:expression}, part (ii)
}
\notag
\\
&
\rho_1 \join \rho_2, e_2' \fwd v' \sqgeq \mu(\ell)
&
\text{
   $\rho_1 \compatible \rho_2$;
   \corref{fwd:monotonicity:expression}
}
\locallabel{assign-premise-two}
\\
&
\rret{v} \sqleq \rret{\tunit}
&
\text{
   inversion on $\eval$
}
\notag
\\
&
\qedLocal
(\update{\update{\mu}{\ell}{\Hole}}{\ell}{v'}, \tunit) 
\sqgeq 
(\mu, v)
&
\text{
   $v' \sqgeq \mu(\ell)$;
   reflexivity
}
\notag
\\
&
\qedLocal
\derivation{\derivationWidth}{
\begin{smathpar}
\inferrule*
{
   \rho_1 \join \rho_2, e_1' \fwd \ell
   \\
   \rho_1 \join \rho_2, e_2' \fwd v'
}
{
   \rho_1 \join \rho_2, 
   \update{\mu}{\ell}{\Hole}, 
   \tassign{e_1'}{e_2'}, 
   \tassignl{e_1'}{\ell}{e_2'}
   \fwd
   \update{\update{\mu}{\ell}{\Hole}}{\ell}{v'}, \rret{\tunit}
}
\end{smathpar}
}
&
\text{
   (\localref{assign-premise-one}, 
    \localref{assign-premise-two}) 
}
\notag
\intertext{\crossrule}
&
\caseDerivation{\derivationWidth}{
\begin{smathpar}
\inferrule*
{
   \mu(\ell) = \Hole
}
{
   \mu, \rret{v}, \tassignl{e_1}{\ell}{e_2} 
   \bwd 
   \Hole, \update{\mu}{\ell}{\Hole}, \tassign{\Hole}{\Hole}, \tassignl{\Hole}{\ell}{\Hole}
}
\end{smathpar}
}
&
\notag
\\
&
\qedLocal
\derivation{\derivationWidth}{
\begin{smathpar}
\inferrule*
{
   \Hole, \Hole \fwd \Hole
}
{
   \Hole, \update{\mu}{\ell}{\Hole}, \tassign{\Hole}{\Hole}, \tassignl{\Hole}{\ell}{\Hole} 
   \fwd
   \update{\update{\mu}{\ell}{\Hole}}{\ell}{\Hole}, \rret{\tunit}
}
\end{smathpar}
}
&
\text{
   \figref{slicing-expression}
}
\notag
\\
&
\qedLocal
(\update{\update{\mu}{\ell}{\Hole}}{\ell}{\Hole}, \rret{\tunit}) 
=
(\mu, \rret{\tunit})
\sqgeq 
(\mu, \rret{v})
&
\text{
   $\mu(\ell) = \Hole$
}
\notag
\intertext{\crossrule}
&
\caseDerivation{\derivationWidth}{
\begin{smathpar}
\inferrule*
{
   \ell, e \bwd \rho, e'
}
{
   \mu, \rret{v}, \tderefl{\ell}{e} 
   \bwd 
   \rho, \mu \join \update{}{\ell}{v}, \tderef{e'}, \tderefl{\ell}{e'}
}
\end{smathpar}
}
&
\notag
\\
&
\rho, e' \fwd \ell \sqgeq \ell
&
\text{
   \thmref{galois-connection:expression}, part (ii)
}
\locallabel{deref-premise}
\\
&
\qedLocal
\mu \join \update{}{\ell}{v} \sqgeq \mu
&
\notag
\\
&
\qedLocal 
\rret{(\mu \join \update{}{\ell}{v})(\ell)} 
=
\rret{(\mu(\ell) \join v)}
\sqgeq 
\rret{v}
&
\notag
\\
&
\qedLocal
\derivation{\derivationWidth}{
\begin{smathpar}
\inferrule*
{
   \rho, e' \fwd \ell
}
{
   \rho, \mu \join \update{}{\ell}{v}, \tderef{e'}, \tderefl{\ell}{e'}
   \fwd 
   \mu \join \update{}{\ell}{v}, \rret{(\mu \join \update{}{\ell}{v})(\ell)}
}
\end{smathpar}
}
&
\text{
   (\localref{deref-premise}) 
}
\notag
\end{flalign}
\end{proof}

\section{Implementation: full example}
\label{app:impl-example}

Below is the full code of Gaussian elimination method example discussed in
section~\ref{sec:implementation}.  As can be seen, our implementation correctly
determines that arrays \texttt{as}, \texttt{bs} and \texttt{bs'} are completely
irrelevant.  Moreover, fragment of code reponsible for zeroing matrix elements
above the diagonal is also correctly marked as irrelevant.

Code of Gauss method has been adapted from
\url{https://rosettacode.org/wiki/Gaussian_elimination#C}

\begin{Verbatim}[commandchars=\\\{\}]
let n = 4 in
let as = \(\slice{[| [|  3.0; -1.0;  2.0; -1.0 |]}\)
         \(\slice{;  [|  1.0;  2.0; -1.0;  2.0 |]}\)
         \(\slice{;  [|  3.0; -1.0;  1.0;  1.0 |]}\)
         \(\slice{;  [| -1.0;  1.0; -2.0; -3.0 |] |]}\) in
let bs = \(\slice{[| -13.0; 21.0; 1.0; -5.0 |]}\) in

-- same system as before but 2nd and 3rd row are now swapped leading to division
-- by 0
let as' = [| [|  3.0; -1.0; \(\slice{2.0}\); \(\slice{-1.0}\) |]
          ;  [|  3.0; -1.0; \(\slice{1.0}\); \(\slice{ 1.0}\) |]
          ;  [|  1.0;  2.0; \(\slice{-1.0}\); \(\slice{ 2.0}\) |]
          ;  [| \(\slice{-1.0}\);  \(\slice{1.0}\); \(\slice{-2.0}\);\(\slice{-3.0}\) |] |] in
let bs' = \(\slice{[| -13.0; 1.0; 21.0; -5.0 |]}\) in

let gauss = fun gauss (a : array(array(double))) (b : array(double)) : array(double) =>
   let dia = ref 0 in
   -- zero elements below the diagonal
   (while !dia < n do
    let row = ref (!dia + 1) in
    (while !row < n do
      let tmp = a[!row][!dia] / a[!dia][!dia] in
      let col = ref (!dia + 1) in
      (while !col < n do
        a[!row][!col] <- a[!row][!col] - (tmp * a[!dia][!col]);;
        col := !col + 1
      ) ;;
      \(\slice{a[!row][!dia] <- 0.0}\) ;;
      \(\slice{b[!row] <- b[!row] - tmp * b[!dia]}\) ;;
      row := !row + 1
    ) ;;
    dia := !dia + 1
   ) ;;
   -- zero elements above the diagonal
   \(\slice{let row = ref (n - 1) in}\)
   \(\slice{let x = array(n, 0.0) in}\)
   \(\slice{(while !row >= 0 do}\)
     \(\slice{let tmp = ref (b[!row]) in}\)
     \(\slice{let j = ref (n - 1) in}\)
     \(\slice{(while !j > !row do}\)
        \(\slice{tmp := !tmp - (x[!j] * a[!row][!j])}\) ;;
        \(\slice{j := !j - 1}\)
     \(\slice{)}\) ;;
     \(\slice{x[!row] <- !tmp / as[!row][!row]}\) ;;
     \(\slice{row := !row - 1}\)
   \(\slice{)}\);; \(\slice{x}\) in
map (fun (a,b) => gauss a \(\slice{b}\)) [ \(\slice{(as, bs)}\) ; (as', \(\slice{bs'}\)) ]
\end{Verbatim}

\fi

\end{document}